\documentclass{aastex}
\usepackage{emulateapj5}

\newcommand{\simgt}{\lower 2pt \hbox{$\, \buildrel {\scriptstyle >}\over {\scriptstyle\sim}\,$}}
\newcommand{\simlt}{\lower 2pt \hbox{$\, \buildrel {\scriptstyle <}\over {\scriptstyle\sim}\,$}}

\newcommand{\um}{UM~425}

\newcommand{\pgone}{PG~1115+080}
\newcommand{\apm}{APM~08279+5255}

\newcommand{\chandra}{{\emph{Chandra}}}

\newcommand{\xmm}{\emph{XMM-Newton}}

\slugcomment{Received 2006 Sep 27; accepted 2006 Dec 24}
\shorttitle{Refined Outflow Constraints of PG~1115+080}
\shortauthors{CHARTAS ET AL.}

\begin{document}

\def\sarc{$^{\prime\prime}\!\!.$}
\def\arcsec{$^{\prime\prime}$}
\def\beginrefer{\section*{References}%
\begin{quotation}\mbox{}\par}
\def\refer#1\par{{\setlength{\parindent}{-\leftmargin}\indent#1\par}}
\def\endrefer{\end{quotation}}

\title{{\sl XMM-Newton} and {\sl Chandra} Spectroscopy of the 
Variable High-Energy Absorption of
PG~1115+080: Refined Outflow Constraints}

\author{G. Chartas,\altaffilmark{1} W. N. Brandt,\altaffilmark{1} S. C. Gallagher,\altaffilmark{2} and D. Proga\altaffilmark{3}}

\altaffiltext{1}{Department of Astronomy \& Astrophysics, Pennsylvania State University,
University Park, PA 16802, chartas@astro.psu.edu, niel@astro.psu.edu}

\altaffiltext{2}{Department of Physics \& Astronomy, University of California -- Los Angeles, Mail Code 154705, 475 Portola Plaza, Los Angeles, CA 90095-1547, sgall@astro.ucla.edu}

\altaffiltext{3} {Department of Physics, University of Nevada, 4505 South Maryland Parkway,
Las Vegas, NV 89154, dproga@physics.unlv.edu}

\begin{abstract}

We present results from multi-epoch spectral analysis of {\sl XMM-Newton} 
and {\sl Chandra} observations of the mini broad absorption line (BAL) quasar PG~1115+080.
This is one of the few X-ray detected mini-BAL quasars to date that is bright 
enough in the X-ray band, mostly due to large gravitational-lensing magnifications,
to allow in-depth spectral analysis.
The present \xmm\ observations of {PG~1115+080} have provided 
the highest signal-to-noise X-ray spectra of a mini-BAL quasar obtained to date.
By modeling the spectra of PG~1115+080 
we have obtained constraints on the column density
and ionization state of its outflowing absorbing gas.
A comparison between these constraints over several epochs
indicates significant variability in 
the properties of the outflowing absorbers in \pgone. 
The depths of the high-energy broad absorption features 
in \pgone\ show a significant decrease 
between the first two observation epochs separated by a rest-frame timescale of $\sim$ 1~year. 
This variability supports the intrinsic nature of these absorbers.
Assuming the interpretation that the 
high-energy absorption features arise from highly ionized Fe~XXV
we constrain the fraction of the total bolometric 
energy released by quasars \pgone\ and \apm\  into the IGM in
the form of kinetic energy to be  
$\epsilon_{\rm k} = 0.64_{-0.40}^{+0.52}$ (68\% confidence), 
and $\epsilon_{\rm k} =0.09_{-0.05}^{+0.07}$, respectively. 
According to recent theoretical studies this range of efficiencies
is large enough to influence significantly the formation of 
the host galaxy and to regulate the growth of the central black hole.
\end{abstract}


\section{INTRODUCTION}
In recent years there has been mounting evidence from both theoretical and 
observational studies for the importance of quasar outflows in regulating the 
growth of supermassive black holes, controlling the formation 
of host galaxies, and enriching the Intergalactic Medium (IGM).
Models of structure formation in a $\Lambda$-CDM cosmology are not
consistent with observations unless feedback, either from star formation
or active galactic nuclei, is included in the simulations.  Recently, the
potential importance of quasar outflows has been explicitly demonstrated
in theoretical models of structure formation and galaxy 
mergers that incorporate the effects of quasar outflows 
[e.g., Scannapieco \& Oh 2004 (SO04); Granato et al. 2004 (G04); 
Springel, Di Matteo, \& Hernquist 2005 (SDH05); Hopkins et al. 2005, 2006]. A basic assumption in the models of 
SO04 is that all quasars host outflows. SO04 find that by choosing the fraction of the total 
bolometric energy released over a quasar's lifetime into the Interstellar Medium (ISM) and IGM
in the form of kinetic energy to be $\epsilon_{\rm k}$=0.05, they can successfully model the observed evolution 
of the ${\it B}$-band quasar luminosity function between redshifts of 0.25 and 6.25. In a recent 
study SDH05 simulated the growth of black holes in 
gas-rich galaxies with and 
without the presence of accretion feedback. The feedback in their model is thought to occur 
through energetic quasar outflows that interact with the gas of the host galaxy. It is assumed that about 5\% 
of the radiated luminosity is thermally coupled via these outflows to the 
surrounding gas. They find that the growth of the black-hole mass is self-regulated 
and eventually saturates at a final value that depends on the initial amount of gas available for accretion. These authors also found that feedback from stars and quasars in their simulations of galaxy mergers can heat and expel gas from the centers of the merged galaxies. This loss of gas during the final stages of the merger can halt nuclear starburst activity and 
cause the merger galaxy to become a gas-poor elliptical.

With the advent of \xmm\ and \chandra\ it has become possible to infer the kinematic and
ionization properties of highly ionized X-ray absorbers in Seyfert 1 galaxies 
(e.g., Kaspi et al. 2002; Kaastra et al. 2002; Netzer et al. 2003). Approximately 60\% 
of Seyfert 1 galaxies show outflowing X-ray and UV absorption by ionized gas with 
velocities up to $\approx$ 2,500 km~s$^{-1}$ (e.g., Crenshaw et al. 1999; Kriss 2002). The absorbing outflow properties of more 
luminous quasars appear to differ in several ways from those of Seyfert 1s. The fraction of occurrence of 
X-ray and UV absorption has been reported to be lower in quasars 
(e.g., Ganguly et al. 2001; George et al. 2002); 
however, the claims for a lower occurrence of X-ray absorption have been challenged 
by Porquet et al. (2004) who find that about half 
of their sample of 21 low-redshift  PG quasars harbor ionized absorbers. 
UV spectroscopic observations indicate that about 20\% 
of quasars show broad absorption lines blueward of their resonant 
UV emission lines (e.g., Hewett \& Foltz 2003). Outflow velocities of the UV absorbers in BAL quasars have been found to 
be as high as \hbox{$\approx$ 60,000~km s$^{-1}$}.

Recent X-ray observations of the mini-BAL quasar \pgone\ and 
BAL quasar \apm\ have suggested the presence of 
relativistic outflows of highly ionized (ionization parameter of the order 
of $\log\xi = 3.5$) absorbing material detected in the iron region ($ >6 $~keV)
with velocities of up to $\sim$0.4$c$ (Chartas et al. 2002, 2003). 
The inferred hydrogen column densities ranging between
10$^{22-23}$~cm$^{-2}$ and relativistic velocities of these outflowing X-ray absorbers imply 
mass-outflow rates that are comparable to the estimated accretion rates.
The presence of massive, highly ionized, and high-velocity outflows from quasars indicates that these winds may be providing significant feedback to the surrounding gas. 
Additional observational evidence to support the presence of quasar feedback came with the detection of high-velocity blueshifted absorption-line features in the X-ray spectra of several quasars and Narrow-Line Seyfert 1 galaxies (Reeves et al. 2003;  Pounds et al. 2003a, 2003b).
We note, however, that some of these claims have been disputed in recent re-analyses of the data 
(e.g., Kaspi \& Behar 2006; McKernan  et al. 2004, 2005).

The relatively small velocity spread observed
for the UV absorption in PG~1115+080, compared to the typical range of 
5,000 -- 25,000~km s$^{-1}$ observed in BAL quasars, suggests that PG~1115+080 be classified 
as a mini-BALQSO (Turnshek 1988; Barlow, Hamann, \& Sargent 1997).
We will consider PG~1115+080 as a mini-BAL throughout this paper 
with the following model in mind. According to the "unification" model for BAL 
quasars most quasars have outflowing winds,
however, BAL and mini-BAL quasars correspond to quasars with
relatively large inclination angles. In particular, in BAL quasars it is commonly thought
that our line of sight intersects a large portion of the outflowing absorber
whereas in mini-BAL quasars our line of sight intersects a
shorter portion. According to this model
mini-BAL quasars would be expected to have smaller absorbing columns, and smaller velocity
gradients than traditional BAL quasars.

In this work we present recent results from monitoring X-ray observations of the mini-BAL quasar 
PG~1115+080 ($z$ = 1.72).
The goal of these observations was to monitor the time variability of the absorption features and thereby 
constrain the kinematic, ionization, and absorption properties of the quasar outflows in this X-ray-bright mini-BAL quasar. Such constraints will allow us to estimate the contribution of quasar winds to the enrichment of the ISM and IGM
and obtain a better understanding of the connection between black-hole and bulge growth in the host galaxy.

 Throughout this paper we adopt a $\Lambda$-dominated cosmology with $H_{0}$ = 70~km~s$^{-1}$~Mpc$^{-1}$, 
$\Omega_{\rm \Lambda}$ = 0.7, and  $\Omega_{\rm M}$ = 0.3.

\section{OBSERVATIONS AND DATA ANALYSIS}

\pgone\ was observed with \xmm\ (Jansen et al. 2001) on 2001 Nov 25, 2004 June 10, 
and 2004 June 26, for 62.9~ks, 81.2~ks and 86.3~ks, respectively.
It was also observed with the Advanced CCD imaging 
Spectrometer (ACIS; Garmire et al. 2003) on board the {\it Chandra X-ray Observatory} 
(hereafter \chandra) on 2000 June 2 and 2000 {Nov~3} for 26.8~ks and 
10.0~ks, respectively.
The spectral analysis of the \chandra\ observations of \pgone\ has been presented in
Gallagher et al. (2002) and Chartas et al. (2003).
Because of recent significant improvements in the calibration of the
instruments on board \chandra\ and \xmm\ since the publication of the
\pgone\ results, we have re-analyzed all observations.
Updates on the calibration of \chandra\ and \xmm\ are reported 
on the \chandra\ X-ray Center (CXC) and \xmm\ Science Operations Centre (SOC) 
World Wide Web (WWW) sites, respectively.\footnote{The WWW sites listing the updates are located at 
\url{http://asc.harvard.edu/ciao/releasenotes/history.html } and \url{http://xmm.vilspa.esa.es/external/xmm\_sw\_cal/calib/rel\_notes/index.shtml},  respectively.}
We analyzed the \xmm\ data for  
\pgone\ with the standard analysis software SAS version 6.1 provided by the \xmm\ SOC.  
The \chandra\ observations of 
\pgone\ were analyzed using the standard software CIAO 3.2 provided by the CXC. 
A log of the observations that includes 
observation dates, observed count rates, 
total exposure times, 
and observational identification numbers is presented in Table 1. We note that the
count rate and total number of counts ($\sim$31,430 counts from the three \xmm\ observations) 
for PG 1115+080 are the highest of any mini-BAL 
quasar X-ray spectrum observed to date.

For the reduction of the \xmm\ observations we filtered the PN (Str{\" u}der et al. 2001) and MOS (Turner et al. 2001)
data by selecting events corresponding to instrument \verb+PATTERNS+
in the 0--4 (single and double pixel events) and 0--12 ranges, respectively.
Several moderate-amplitude background flares were present
during the \xmm\ observations of \um\ and \pgone.
The PN and MOS data were filtered to exclude times when the full-field count rates
exceeded 20~cnts~s$^{-1}$ and 4~cnts~s$^{-1}$, respectively.   
The extracted spectra from the PN and MOS 
were grouped to obtain a minimum of 100 and 40 counts, respectively, in each energy bin,
allowing use of $ \chi^{2}$ statistics.
Background spectra for the PN and MOS detectors were extracted
from source-free regions near \um\ and \pgone.
The PN and MOS spectra were then fitted 
simultaneously with a variety of models employing 
\verb+XSPEC+ version 11.3 (Arnaud 1996).  The energy ranges used for fitting the PN and MOS 
data were 0.2--10~keV and 0.4--9~keV, respectively.

For the reduction of the \chandra\ observations we used 
standard CXC threads to screen the data for 
status, grade, and time intervals of acceptable aspect solution and background levels.
The pointings of the observatory placed 
\pgone\ on the back-illuminated S3 chip of ACIS.
To improve the spatial resolution we
removed a {$\pm$~0\sarc25} randomization applied to the event positions
in the CXC processing and employed a sub-pixel resolution technique
developed by Tsunemi et al. (2001).

In both the \xmm\ and \chandra\ analyses we tested the sensitivity of our results 
to the selected background and source-extraction
regions by varying the locations of the background regions and varying the 
sizes of the source-extraction regions. We did not find any significant change in the 
background-subtracted spectra.  For all models of  
\pgone\ we included Galactic absorption due to neutral gas with 
a column density of  
$N_{\rm H}$=3.5 $\times$ 10$^{20}$~cm$^{-2}$(Stark et al. 1992). 
All quoted errors are at the 90\% confidence level unless mentioned otherwise
with all parameters conservatively taken to be of interest except absolute normalization.

\subsection{Spectral Analysis of the \xmm\ Observations of \pgone.}
We first simultaneously fitted the PN, MOS1 and MOS2  
spectra of \pgone\ at each of the three epochs with a model consisting of a power law with neutral intrinsic
absorption at $z = 1.72$ (model 1 of Table 2). 
These fits support the presence of an intrinsic absorber with
column densities of  $N_{\rm H}$= 2.2$_{-0.5}^{+0.5}$ $\times$ 10$^{21}$~cm$^{-2}$,
4.6$_{-0.5}^{+0.5}$ $\times$ 10$^{21}$~cm$^{-2}$, and 
2.9$_{-0.7}^{+0.7}$ $\times$ 10$^{21}$~cm$^{-2}$, for 
the three \xmm\ observations of \pgone.
The fits are not acceptable in a statistical sense (see the reduced $\chi^{2}$ values in Table 2). 
The fit residuals show significant absorption at observed-frame energies of 2.5--5~keV.
To illustrate the presence of these features and absorption features
below 0.6~keV, we fit the spectra from observed-frame 2--2.5~keV and 
5--10~keV with a power-law model
(modified by Galactic absorption) and extrapolated this model to the energy ranges not fit
(see Figure 1). 
For clarity we only show the higher S/N ratio PN data in Figure~1; however, 
all fits were performed simultaneously using the PN and MOS1+2 data unless mentioned otherwise.
The lower panels in Figure 1 show the 
$\Delta\chi$ residuals between the best-fit power-law model and the PN data.
We note that the apparent variability in the amplitude of the absorption features 
is not due to variations in the best-fit value of $\Gamma$ 
for the spectral fits performed within the observed-frame ranges of 2--2.5~keV and 
5--10~keV. The best-fit values of the photon 
indices for epochs 1, 2 and 3 
were $1.91_{-0.10}^{+0.10}$, $1.88_{-0.10}^{+0.10}$,
and $1.84_{-0.11}^{+0.11}$.
For the purpose of comparing the absorption residuals 
between epochs the photon indices for epochs 1, 2, and 3 were set to $\Gamma$ = 1.9.
We proceed in fitting a variety of models to the data guided by the shape and location of these identified 
absorption residuals.  
As our first refinement we considered an absorbed power-law model with  
Gaussian absorption lines
near the absorption features appearing between observed-frame energies of ~2 and 5~keV (see model 2 of Table 2).
Hereafter, we refer to these absorption features as $abs1$ and $abs2$. 
The fits to the spectra of \pgone\ for epochs 1 and 2 contained two absorption lines, and the fits 
for epoch 3 contained one absorption line. The inclusion of these 
absorption lines in model~2 resulted in significant 
improvements of the fits compared to the previous ones in model 1  
at the $ > $ 99.9\%, 99.2\% and 93.6\% confidence levels (according to the $F$-test)
for epochs 1, 2 and 3, respectively. 

The remaining most significant contributions to the large values of $\chi^{2}$ for model 2 
arise from the residuals below 0.6~keV.
To model these residuals we replaced the neutral absorber in our spectral model
with an ionized intrinsic absorber (see model 3 of Table 2).
These low-energy residuals are commonly detected in moderate 
S/N spectra of BAL quasars and are thought to
arise from absorption by multiple ions of O, Ne, Na, Mg, and Fe and/or partial covering. 
In particular, we used the \verb+absori+ model contained in 
\verb+XSPEC+ (Done et al. 1992).
We note that the \verb+absori+ model is just
a first approximation to what is likely a more complex situation.
The temperature of the ionized absorber is not calculated 
self-consistently by proper thermal balance in 
the \verb+absori+ model but considered as an input parameter.
We found a significant improvement in fit quality at the $ > $ 99.9\% 
confidence level according to the $F$-test (see model 3 of Table 2). 
The best-fit rest-frame energies and widths  
of the absorption features for model 3 for epochs 1, 2 and 3 are listed in Table 2.
We note that the best-fit energies of these features are quite stable
when trying different models for the low-energy spectral complexity.
As an independent check of the accuracy of the fits that used the \verb+absori+ model we also
repeated several of these fits using the warm absorber model \verb+XSTAR+.
\verb+XSTAR+ calculates the physical conditions and emission spectra of photoionized gases.
In the current analysis we use a recent implementation of  the \verb+XSTAR+ model that can be used within
\verb+XSPEC+. 

We found that the best-fit parameters of the low-energy ionized absorber derived using the \verb+XSTAR+ model
were consistent with the values found using the \verb+absori+ model.
We performed fits to the spectra of \pgone\ at the three epochs using both \verb+absori+ and \verb+XSTAR+ 
to compare results and performed the estimates of the confidence contours of
the best-fit parameters using the more time-efficient \verb+absori+ model.

We investigated whether the observed flux variability of the high-energy absorption features was consistent 
with a decrease in the strength of two absorption lines used to model the absorption features. 
The energies and widths of the absorption lines were held fixed between epochs.
Specifically, we performed a simultaneous fit to all epochs
incorporating model 2 of Table 2 with the difference that
the energies and widths of the absorption lines were 
constrained to be the same and only the normalizations
of the absorption lines were allowed to vary. These fits were performed in the 1--10~keV band of the higher S/N PN data.
We obtained a $\chi^{2}$ of 137.1 for 137 dof, and the best-fit rest-frame energies of 
the lines were 7.5 $\pm$ 0.3~keV and 10.0 $\pm$ 0.4~keV. 
Allowing the energies and widths of the lines to vary resulted in a similar $\chi^{2}$ of 132.0 for 129 dof. 
We note that for the fits where the energies and widths were constrained to be the same between epochs
we found that the best-fit values of the optical depths for the absorption lines at 7.5~keV 
were consistent with zero for epochs 2 and 3. This was expected since an inspection of Figure 1 indicates that 
there are no residuals for epochs 2 and 3 near the rest-frame energy of 7.5~keV (left vertical dashed line) .
We conclude that the strengths of the absorption lines for epochs 2 and 3 are too weak
to infer any variability of their energies or widths.

We next replaced the ionized absorber with a partial-covering model and repeated the fits for each epoch.
We find that fits that include partial-covering (see model 4 of Table 2) provide $\chi^{2}$ 
values and quality-of-fit parameters comparable to
those that included an ionized absorber. 
Models that included absorption edges for the high-energy absorption features of \pgone\ for \hbox{epoch 1}
were considered and rejected in the current analysis, that includes 
the new calibration, as previously found in Chartas et al. (2003).
The basic problem with such edge models is that, for plausible iron abundances, they predict too
much absorption at low energies.
The inclusion of a broad (width allowed to vary in fits) Fe K$\alpha$ emission line in all models in Table 2 for epoch 1
resulted in larger reduced $\chi^{2}$ values compared to those with no line included;
therefore, there is no evidence for statistically significant Fe emission.
We also tested the sensitivity of the best-fit parameters to the energy ranges used for fitting
the PN data by varying the low-energy boundary of the fits from 0.2~keV to 0.4~keV.
We did not find any significant change in the parameters of the high-energy absorption
lines. The exclusion of the 0.2--0.4~keV PN data, however, did result in slightly poorer
constraints on the ionization parameter of the gas causing the low-energy absorption. 
A detailed description of the current calibration status of the \xmm\ instruments
can be found on the \xmm\ SOC WWW site.  
A recent energy-dependent re-working of the PN 
response has resulted in a significant improvement in the calibration down to
0.2~keV and brings the PN and MOS detectors into better agreement.

In Figure 2a we show the 68\% and 90\% confidence contours (based on model 3 of Table~2) for the
photon indices versus column densities for the three epochs.
The column densities for the low-energy absorbers in models 3 and 4 are
consistent with no variation.
A possible weak variation of the photon index between epochs 1 and 3 is detected.
68\% and 90\% confidence contours for the ionization parameter of the low-energy absorber 
versus column density are presented in Figure 2b. The absorber in \pgone\ appears to have been 
more ionized during epoch 1. In particular, during epoch 1 the 
ionized absorber is characterized by an ionization parameter of
$\xi = L/nr^{2} = 145_{-52}^{+84}$~erg~cm~s$^{-1}$ and a hydrogen column density
of $N_{\rm H} = 1.33_{-0.35}^{+0.58}$ $\times$ 10$^{22}$~cm$^{-2}$,
and during epoch 2 by $\xi =  56_{-23}^{+30}$~erg~cm~s$^{-1}$ and a hydrogen column density
of $N_{\rm H} = 1.41_{-0.27}^{+0.36}$ $\times$ 10$^{22}$~cm$^{-2}$,
where $L$ is the integrated 5~eV--300~keV incident luminosity, $n$ is the
electron number density of the absorber, and $r$ is the distance between the
absorber and ionizing source. 

Figure 2 suggests that the ionization parameter
decreased in the order of epochs 1, 3, and 2.
This trend in ionization parameter is evident in
the PN data shown in Figure 1.
For epoch 1, as shown in Figure 1a, the spectrum at low energies $\sim$ 0.2--0.4~keV 
appears less absorbed relative to epochs 2 and 3, 
as expected for a more highly ionized absorber where the
opacity from low-z metals is smaller.
Additional confirmation for the trend in ionization parameter is 
provided in \S3.1.

Our spectral analysis of \pgone\ implies the existence of two distinct absorbers
with very different ionization parameters.
In particular, if we interpret the apparent detection of absorption
lines at rest-frame energies of 7.27$_{-0.10}^{+0.37}$~keV 
and 9.79$_{-1.05}^{+0.96}$~keV as being due to Fe, the most
conservative assignment (giving the lowest outflowing velocity) is to
highly ionized Fe XXV which requires $\log\xi \sim 3.5$.
In contrast, the significant absorption features below rest-frame energies of 
$\sim$ 1.6~keV are best fit with an ionization parameter of the order of $\log\xi = 2$ 
(see Table 2). A justification of the interpretation of the absorption lines
 is given in \S 3 and presented in more detail in 
Chartas et al. (2002, 2003).
To illustrate better the absorption lines that can
arise at these two distinct ionization levels
we show in Figure 3 the expected absorption spectrum of \pgone\ in the ${1-10~{\rm \AA}}$ 
rest-frame range as derived with the photoionization code XSTAR 
assuming a slab-like stationary absorber with
 $N_{\rm H}$ = 9 $\times$ 10$^{22}$~cm$^{-2}$, $\log\xi = 2$, and $\log\xi = 3.5$.
The assumed column density of 9 $\times$ 10$^{22}$~cm$^{-2}$ lies between the estimated values of the column densities of the high ionzation absorbers of PG~1115+080 listed in Table 4.
At ionization levels of $\log\xi \sim 3.5$ and rest-frame wavelengths in the range 
1.5--2~{\AA} (see panel a of Figure 3) the Fe lines are quite isolated 
and are therefore apparent even with low-resolution spectroscopy. 
At rest-frame wavelengths in the range of 6.2--10~{\AA} and at ionization parameters of 
$\log\xi \sim 2$  the spectra contain a larger density of strong absorption lines that are difficult to resolve and identify (see panel b of Figure 3).

The derived column densities for the low-ionization absorber of PG~1115+080 
(see Table 2) for the three epochs described in this paper
are lower than those found in X-ray observations of typical BAL quasars.
For example, the column densities of BAL quasars as inferred from 
X-ray observations lie in the range of 0.4--250 $\times$ 10$^{23}$ cm$^{-2}$ 
(e.g., Gallagher et al. 2002 and Punsly 2006 present BAL-quasar column densities derived from X-ray observations).
We emphasize that most current column-density estimates derived from 
X-ray observations are poorly constrained due to the limited S/N of most 
existing X-ray spectra of BAL and mini-BAL quasars.
The presence of ionized or partially covering absorption cannot be
accurately inferred from the analysis of individual X-ray spectra of most BAL quasars
and if present would in general lead to an underestimate of the column densities.
On the other hand, the large lensing magnifications
of APM~08279+5255 and PG~1115+080 have provided moderate S/N X-ray
spectra for these objects resulting in improved constraints
compared to unlensed quasars.
We note that the column density of PG~1115+080 appears to have varied significantly since an
earlier observation. Specifically, during a {\it ROSAT} PSPC observation of
PG~1115+080 on 21 November 1991 the 0.2--2~keV X-ray flux of
this object was found to be at a minimum level (low-state) of about 
0.4 $\times$ 10$^{-13}$~erg~s$^{-1}$~cm$^{-2}$
(see Figure 6 of Chartas 2000) and the
intrinsic column was constrained
to be 1.2 $\pm$ 1.1 $\times$ 10$^{23}$~cm$^{-2}$. This column density is consistent with the typical values observed in BAL quasars.
We estimated the optical-to-X-ray spectral slope, quantified 
as \( \alpha_{\rm ox} = \log(f_{\rm 2~keV}/f_{2500 {\rm \AA}})/\log(\nu_{\rm 2~keV}/\nu_{2500 {\rm \AA}}) \) 
(Tananbaum, et al. 1979), 
where $f_{\rm 2~keV}$  and $f_{2500 {\rm \AA}}$ are the flux densities 
at 2~keV and 2500~{\AA} in the quasar rest-frame, respectively.
For the cases of unabsorbed(absorbed) flux densities at 2~keV for the epoch 1
{\it XMM-Newton} observation of PG~1115+080 we find $\alpha_{\rm ox}$ = -1.63(-1.67). 
We note that from the absorbed flux 
density at 2~keV for the 1991 {\it ROSAT} PSPC observation (low-state)
of PG~1115+080 we find $\alpha_{\rm ox}$ = -1.97. 
Due to the limited quality of the {\it ROSAT} PSPC spectrum 
useful constraints on the unabsorbed value of $\alpha_{\rm ox}$ 
could not be obtained for the 1991 observation.
For comparison the value of $\alpha_{ox}$ 
expected for a typical AGN with a 2500~{\AA} luminosity density similar 
to that of the epoch 1 observation of PG~1115+080
based on the recent empirical relation of Steffen et al. 2006 
(see equation 2 of Steffen et al.) is $\alpha_{\rm ox}$ = -1.58. 
The rms of $\alpha_{\rm ox}$ at $\log(L_{2500 {\rm \AA}})$ = 30.8
is about 0.2 (see Table 5 of Steffen et al.).
We conclude that PG~1115+080 is slightly X-ray weak during the three observation epochs
presented in this paper but still within the scatter 
of $\alpha_{\rm ox}$ detected in typical AGN.

\section{DISCUSSION}
\begin{sloppypar}
Our analyses of the \chandra\ and \xmm\ spectra of \pgone\ and \apm\
presented here and in the discovery papers of Chartas et al. (2002, 2003) 
indicate that the high-energy broad absorption features 
detected in these objects
are significant at the $  > $ 99.9\% confidence level.
Our claimed detections of relativistic winds in the X-ray spectra of \pgone\ and \apm\ were based  
on the following steps:

a) We chose a standard approach to fitting the X-ray spectra of \pgone\ and \apm.
In particular, we first started with a simple model and added components
of increasing complexity motivated both by
the residuals of the spectral fits and our understanding of the 
physics and structure of mini-BAL and BAL quasars. 
The first model consisted of a simple power-law
modified by intrinsic absorption. This choice was inspired by numerous
empirical and theoretical studies that indicate that the main X-ray continuum emission component of 
quasars  (e.g., rest-frame 2--30~keV)  
is produced by inverse Compton scattering of soft photons from the accretion disk by 
hot electrons in a corona (e.g., Haardt \& Maraschi 1991; Reeves \& Turner 2000).
An inspection of the residuals to fits with simple absorbed power-law models
indicated the presence of significant complex absorption residuals at
rest-frame energies above $\sim$6.4~keV
and rest-frame energies below $\sim$1.5~keV. 

b)  To account for these fit residuals we considered a variety of spectral models.
Several of these models were rejected in the discovery papers and this work.
The list of rejected models includes the following:
Broken power-law models with and without a soft blackbody component,
models that included one and two absorption edges to account for the high-energy absorption,
models that included neutral absorption at the source to account for the low-energy absorption,
and absorption by possible intervening material to account for the high-energy absorption.
Single and double partial-covering models that try to fit
the high-energy features of \pgone\ with iron edges 
require a curvature of the low-energy continuum that
is inconsistent with our data. Additional data would be required to rule out even more complex 
partial-covering models. 

c) Motivated by the structure of the low and high-energy residuals, the fact that \pgone\
and \apm\ exhibit broad absorption lines in the UV
(e.g., the properties of the UV broad absorption lines in \pgone\ and \apm\ have been presented in 
Michalitsianos et al. (1996) and Irwin et al. (1998), respectively), 
and previous claims of detections of low-energy ionized absorption in  
the X-ray spectra of BAL quasars, we fitted the spectra with a model that consisted of
a power-law modified by ionized absorption at the source and
two iron lines at the source to account for the high-energy absorption.
This model produced acceptable fits to the X-ray spectra of \pgone\ and 
\apm\ and resulted in a significant improvement over all other models considered.
We note that models that included high-energy absorption
lines in \pgone\ and \apm\ resulted in improvements in the 
fits compared to models that did not include absorption lines at the greater than 99.9\% confidence level.
In the present work we also confirm this significant detection by re-analyzing the same data with 
updated instrumental calibration files.

d) We made a plausibility argument based on the energies of the identified lines,
and the known energies of absorption lines from all the abundant elements, that the 
detected absorption lines are associated with highly ionized Fe K$\alpha$ absorption. 
Based on this identification we estimated the velocities of the outflowing X-ray absorbing material.

Several of the key derivations that follow in this section
assume our interpretation that the high-energy absorption
is due to lines arising from highly ionized Fe XXV and/or Fe XXVI
and that the outflow velocities of these absorbers range between 0.1$c$ and 0.4$c$.
Here we provide a summary of our justification of this interpretation
that was presented in more detail in Chartas et al. (2002, 2003).
\end{sloppypar}
Our analyses of the \chandra\ spectrum of \apm\ and the
{\it XMM-Newton} spectrum of \pgone\ showed strong evidence for
the presence of absorption lines at rest-frame energies of 8.05$_{-0.08}^{+0.10}$~keV and
9.79$_{-0.19}^{+0.20}$~keV for \apm\
and rest-frame energies of $7.27_{-0.10}^{+0.37}$~keV and $9.79_{-1.05}^{+0.96}$~keV for \pgone.
For \apm\ we assumed that the 
two lines are produced by resonant absorption due to
Fe~{\sc xxv} 1s--2p (Fe~XXVI~Ly$\alpha$), and inferred that the X-ray absorbers are outflowing with
velocities of $\sim$~$0.20 c$(0.15$c$) and $\sim$~$0.4c$(0.36$c$), respectively.
For \pgone\ we assumed that the
high energy absorption is due to 
two lines produced by resonant absorption due to
Fe~{\sc xxv} 1s--2p (Fe~XXVI~Ly$\alpha$), and inferred that the X-ray absorbers are outflowing with
velocities of $\sim$~$0.1c$(0.05$c$) and $\sim$~$0.4c$(0.36$c$), respectively.
The wind geometry that we assume is based on the unified BAL model 
(e.g., Weymann et al. 1991; Murray et al. 1995; Proga 2000; Elvis 2000).
In this model the AGN outflow originates in the accretion disk
from a narrow range of radii.
This wind rises initially almost perpendicular to the accretion disk
and becomes more radial and equatorial at larger radii to form a bi-cone.
The unified BAL model proposes that most of the observed range of absorption line-widths can be explained
with orientation and with a velocity gradient in the outflowing stream.
The hydrodynamical models of Proga et al. (2000) indicate that the density of the wind
peaks near an inclination angle of about 70 degrees and most of the outflow is
confined within $\pm$ 10 degrees of this angle.
We note that Proga \& Kallman (2004) 
showed that the wind inclination
angle can be smaller and the wind does not need to be equatorial
if it is mostly driven by UV radiation from the accretion disk.

For the velocity calculations we considered the special-relativistic 
velocity correction and assumed that the angle $\theta$ between the wind velocity and 
our line of sight is 20$^{\circ}$. This angle is not constrained with
the present data; however, hydrodynamical simulations indicate that the BAL wind
divergence angle may range between 10$^{\circ}$--30$^{\circ}$
depending on the location of the inner radius of the disk. 
The velocity estimates are not sensitive to this range of angles.
For example, the inferred outflow velocity (assuming $\theta$ $\sim$ 20$^{\circ}$) 
for the 9.79~keV absorption line detected in \pgone\
would vary from $-$6\% to $+$13\% for $\theta$ ranging between 10$^{\circ}$--30$^{\circ}$.

Of all the abundant elements, iron absorption lines would be
the closest in energy to the observed features.
As shown in Figure 3 the strongest lines with rest-frame energies indicated in parenthesis
near the observed ones are:
Fe XXV 1s--2p (6.7~keV), Fe XXV 1s--3p (7.88~keV), 
Fe~XXVI~Ly$\alpha$ (6.97~keV), Fe~XXVI 1s--3p (8.25~keV),
S XV~1s--2p (2.46~keV), S~XVI~Ly$\alpha$ (2.62~keV)
Si XIV Ly$\alpha$ (2.005~keV), Ar XVIII Ly$\alpha$ (3.321~keV),
and Ca XX Ly$\alpha$ (4.104~keV)
(based on the energies of permitted resonance lines of Verner et al. 1996).
In this sense, our interpretation that the absorption lines are associated
with highly ionized Fe~K absorption is the most conservative one possible
(e.g., absorption lines from relativistic sulfur or oxygen would require
much larger blueshifts).
We also investigated  whether intervening absorbers in the lens galaxies
or in possible damped Lyman alpha systems
along the line of sight could explain the high-energy absorption features.
We concluded that the
observed high-energy absorption features
cannot be produced by absorption in intervening
systems and the most likely
origin is intrinsic absorption by highly ionized iron.
The intrinsic origin of the absorption
has been confirmed with the detection of significant flux variability
of the absorption features in \apm\ (see Figure 7 in Chartas et al. 2003)
and now with the detection of significant variability in \pgone\ (see Figure 4)
as discussed in more detail in \S 3.1. 
From the present data we can infer significant variability of the 
normalization of the absorption lines, however, higher quality spectra
will be required to distinguish if there is any energy variability of the absorption lines.
  
Assuming this interpretation for the relativistic X-ray absorbing material, we
present in the following sections estimates of the mass-outflow rate and 
efficiency of the outflow in \pgone\ and \apm.
These estimates are important for addressing the fundamental issue of whether quasar outflows are capable of
influencing their host galaxies and black-hole growth.
Since the present observations of \pgone\ 
do not resolve the lensed images we also estimate the influence of this effect
on our main conclusions.

\subsection{Variability of the outflow in \pgone}
Our spectral analysis of the \xmm\ observations of \pgone\ presented in \S 2.1 indicates 
possible variability of the X-ray broad absorption features between different epochs. 
To quantify the significance of these variations we 
determined the sum of the residuals of $\Delta\chi$ between the best-fit 2--2.5~keV and 5--10~keV 
continuum model and data in the energy range 2.5--5.0~keV
based on the spectral fits presented in Figure 1. 
As described in \S 2.2 we fit the spectra within the energy ranges of 2--2.5~keV and 5--10~keV  
with Galactic absorption and a power-law model
and extrapolated this model to the energy ranges not fit (see \S 2.1).
For this summed-$\Delta\chi$ method to be statistically valid, the best-fit 2--2.5~keV and 5--10~keV models
and the S/N of the spectra need to be similar for all epochs. 
The best-fit 2--2.5~keV and 5--10~keV spectral slopes for all three epochs have essentially the same value of 
$\Gamma = 1.9$ for epochs 1, 2, and 3 respectively, and the S/N of the three observations are similar with 
effective exposure times and source count rates listed in Table 1.
We therefore do not expect variations of the best-fit models 
or S/N of the spectra to contribute strongly to the variations in the high-energy residuals.
We find total $\Delta\chi$ residuals 
of $\Delta \chi_{\rm E1}$ = \(-\)40 $\pm$ 3,  $\Delta \chi_{\rm E2}$ = \(-\)26 $\pm$ 4,
and $\Delta \chi_{\rm E3}$ = \(-\)20 $\pm$ 3 for epochs 1, 2, and 3, respectively.

A second approach to quantifying variability of the high-energy 
absorption features relies on taking ratios of the spectra.
This approach is independent of model assumptions and takes into 
account the uncertainties in each bin.
For the purpose of this analysis the grouping of the spectra was made identical for the three epochs.
In Figure 4 we show the ratios of the spectra of epoch 1 to epoch 2 ($R_{12}$), epoch 1 to epoch 3 ($R_{13}$), 
and epoch 2 to epoch 3 ($R_{23}$). 
We find the mean of the ratios within the high-energy absorbed range 2.5--5.0~keV to be
$ < R_{12} > $ = 0.83 $\pm$ 0.05,  $ < R_{13} > $ = 0.73 $\pm$ 0.05, and $ < R_{23} > $ = 0.88 $\pm$ 0.05.

As a third method of estimating the variability of the X-ray broad absorption features we
compared the strengths of the best-fit Gaussian lines for the three epochs.
Specifically, we found the total absorbed photon fluxes in the Gaussian absorption lines
for epochs 1, 2, and 3 were (2.4 $\pm$ 1.1) $\times$ 10$^{-5}$ photons~cm$^{-2}$~s$^{-1}$,
(0.49 $\pm$ 0.20) $\times$ 10$^{-6}$ photons~cm$^{-2}$~s$^{-1}$, 
and (0.34 $\pm$ 0.16) $\times$ 10$^{-6}$ photons~cm$^{-2}$~s$^{-1}$, respectively.
The decrease in the total absorbed photon fluxes in the absorption lines is consistent with the results of the other two methods.

All approaches imply significant variability of the X-ray broad absorption features in \pgone\
between epochs 1 and 2 separated by 0.92~yr (rest-frame) and marginal 
variability of the X-ray broad absorption features between epochs 
2 and 3 separated by 5.9~d (rest-frame).
The  5.9~d variability, if real, is consistent with a relatively small launching radius of the 
X-ray absorber. Specifically, assuming a radiatively driven wind 
\footnote{
We find the bolometric luminosity of PG~1115+080 to be
$L_{\rm Bol}$ = 3.3~$\times$~10$^{46}$~erg~s$^{-1}$.
To estimate $L_{\rm Bol}$ we used the empirical relation
$L_{\rm Bol}$ = $f_{\rm BC}(3000 \rm \AA)$$\nu$ $L_{\rm \nu}(3000 \rm \AA){\mu}^{-1}$,
where $f_{\rm BC}(3000 \rm \AA) = 5.3$ is the luminosity-dependent bolometric correction
obtained from equation (21) of Marconi et al. (2004) and
$\mu$ is the flux magnification assumed to be $\sim$ 25
based on lensing models of this system (e.g., Impey et al. 1998).
The value of $L_{\rm \nu}(3000 \rm \AA)$ = 1.57 $\times$ 10$^{32}$ erg~s$^{-1}$~Hz$^{-1}$
for PG~1115+080 was taken from Neugebauer et al. (1987).}
and using equation 1 of Chartas et al. (2003) we estimate that a launching radius of about 
$7~R_{\rm s}$
is needed for the absorber to reach a terminal velocity of about 0.4~$c$.
At a launching radius of $7~R_{\rm s}$
the time to reach 90\% of the terminal velocity is about 9~d. This time-scale is consistent with the marginal detection of variability of the Fe absorption lines between epochs 2 and 3.

A model-independent confirmation of the trend of the 
ionization parameter presented in \S 2.1 can also be seen in Figure 4.
In particular, Figure 4 shows significant positive residuals of the 
ratios $R_{12}$ and $R_{13}$ and negative residuals of the ratio $R_{23}$
within the 0.2--0.4 keV band.
This indicates that the 0.2--0.4~keV flux decreased in the order of 
epochs 1, 3, and 2.
This trend in soft X-ray flux is consistent with the 
observed trends of the 0.2--0.4~keV residuals shown in Figure 1
and the observed trend in ionization parameter 
derived from our spectral analysis and shown in Figure 2. 
We note that the ratio of the EPIC PN spectra is
not sensitive to the precise value of the calibrated effective area 
(the effective area cancels out in the ratio)
as long as the response of the PN detector remained the same between observations.
The PN detector response is known to be stable with time
according to the calibration status of the \xmm\ instruments (see footnote 4).

Significant variability of PG~1115+080 has also been 
reported in the optical, UV and X-ray bands.
Weymann et al. (1980) report several absorption lines as being visible in
the May 1980 observations of this quasar.
Young et al. (1982) detect only weak absorption
features bluewards of the CIV line in their December 1980 spectrum of PG~1115+080
and do not detect the lines seen in the May 1980 observations.
Strong and rapid variability of broad absorption lines blueward of OVI in PG~1115+080
were reported by Michalitsianos et al. (1996).
Chartas et al. (2000) reported a decrease by a factor of about 13 of the 0.2--2~keV flux
between 1979 December 5 and 1991 November 21 and an increase by a factor of about
5 between the 1991 November 21 and 1994 May 27 observations of PG 1115+080.

Several recent theoretical studies of high-velocity outflows from quasars 
(e.g., Everett \& Ballantyne 2004; Sim et al. 2005)  assume that
the high-ionization absorption lines of S~XIV, S~XVI and Fe~XXV observed in several quasars  
(i.e., PG 1211+143 and PG 0844+349 
reported by Pounds at al. 2003a,b) 
originate from gas that has reached its terminal velocity. This terminal velocity is then approximated 
with the escape velocity from the region from which the wind is launched, resulting in the approximation
$R_{\rm launch} \sim R_{\rm s}(c/v_{\rm obs})^{2}$, where $v_{\rm obs}$ is the observed outflow velocity and $R_{\rm s}$
is the Schwarzschild radius. For $v_{\rm obs}$ in the range 0.1--0.4$c$ the expected launching radii for \pgone\ will lie in the range \hbox{100 -- 6$R_{\rm s}$}. 
For an estimate of the black-hole mass we used the empirical relation (equation 7)
of Vestergaard \& Peterson (2006).
Specifically, we find the mass of the black hole and the Schwarzschild radius
to be $M_{\rm bh}$ = 1.1 $\times$ 10$^{9}$ M$_{\odot}$
and $R_{\rm s}$ = 2G$M_{\rm bh}$/$c^{2}$ = 3.1 $\times$ 10$^{14}$~cm, respectively,
based on the FWHM of CIV of 4700 km~s$^{-1}$ taken from the Hale spectrum of
PG~1115+080 (Young et al. 1982), and the 1350 ${\rm \AA}$ luminosity density
of $L_{\nu}$ = 9.48 $\times$ 10$^{31}$~${\mu}^{-1}$~erg~s$^{-1}$~Hz$^{-1}$ 
taken from Neugebauer et al. (1987), where $\mu$ is the flux magnification assumed 
to be $\sim$ 25 based on lensing models of this system (e.g., Impey et al. 1998).

%
%
One of the assumptions in these recent theoretical studies is that the absorbers are 
observed near their terminal velocities.  
As we proposed in our previous study of \apm\ another possibility is that we are observing the absorber as it is
being accelerated near the launching radius. We speculate that initially the absorbing 
material supplied by the accretion disk 
has a relatively low ionization parameter because of the large gas density at the base of the wind.
A low ionization parameter results in a large value of the force multiplier that can lead to 
significant acceleration through scattering in atomic resonance lines (line-driving; the force multiplier represents the ratio by which the line opacity of the absorbing material increases the radiation force relative to that produced by Thomson scattering alone).
Because of the intense radiation field of the central source the absorbing gas becomes increasingly ionized
within a time that is much shorter than the time it takes for the absorber to reach its terminal velocity.
When the ionization parameter reaches values of about $\xi$ $\simgt$ 1000 erg~cm~s$^{-1}$ 
absorption from highly ionized lines such as Si~XIV, S~XVI and Fe~XXV
are expected to be produced. As the ionization parameter increases the absorber becomes completely ionized and
line-driving becomes no longer important. This interpretation predicts rapid changes
(compared to those observed in UV broad absorption lines) in 
the observed energies and equivalent widths of the X-ray BALs.  
We note that magnetically driven outflows is also a possibility,
as this might be needed to reach such high velocities if the material
becomes highly ionized quickly.

The average continuum flux of \pgone\ has not varied by more than $\sim$ 20\%  
between epochs as indicated by the observed PN count rates listed in Table 1.
The fractional change of the continuum flux can also be seen in the ratio plots shown in Figure 4.
There is thus no correlation between the depth of the high-energy 
broad absorption features and continuum flux level.

 \subsection{Efficiency of the quasar outflow in \pgone.}
 
One of the key and highly uncertain parameters used in recent theoretical models 
that describe black-hole growth and structure formation
is the efficiency of the quasar outflow. The efficiency is defined as 
the fraction of the total bolometric energy released over a quasar's lifetime into the ISM and IGM 
in the form of kinetic energy injection and can be expressed as:
 
\begin{equation}
\epsilon_{\rm k,i} = {{1}\over{2}}{\dot{M}_{\rm i}{v^{2}_{\rm wind,i}}\over{L_{\rm Bol}}} = 2 {\pi}{f_{\rm c,i}}R_{\rm i}(R_{\rm i}/{\Delta}{R_{i}})N_{\rm H,i}m_{\rm p}{{v^{3}_{\rm wind,i}}\over{L_{\rm Bol}}}
\end{equation}

\noindent
where $\dot{M}_{\rm i}$ is the mass-outflow rate of component $i$, $v_{\rm wind,i}$ is the outflow velocity of the X-ray absorber of component $i$, $f_{\rm c,i}$
is the global covering fraction of the absorber of component $i$, ${\Delta}{R_{\rm i}}$ is the thickness of the absorber at radius $R_{\rm i}$ of component $i$, $N_{\rm H,i}$ is the hydrogen column density of component $i$,
and $L_{\rm Bol}$ is the bolometric photon luminosity of the quasar.
We focus on estimating $\epsilon_{\rm k}$ for the outflow in \pgone\ observed during epoch 1. 
The velocities of the different outflowing X-ray absorbers are listed in {Table~3}.
We assume a conservatively wide range for the covering factor of $f_{\rm c}$=0.1--0.3 
based on the observed fraction of BAL quasars (e.g., Hewett \& Foltz 2003)
and a fraction $R/{\Delta}{R}$ ranging from 1--10 based on current theoretical models
of quasar outflows (e.g., Proga \& Kallman 2000). 
As we argued in $\S$ 3.1 we are likely observing the absorbers as they 
are accelerated near their launching radii. We have assumed absorber 
radii ranging from 3--15 $R_{\rm s}$ based on the maximum observed
X-ray velocity components. 
We used a Monte Carlo approach to estimate the errors ofÊ
$\dot{M}_{\rm i}$ and $\epsilon_{\rm k}$.
The values of $f_{\rm c}$, $R/{\Delta}{R}$, and $R_{\rm i}$ were assumed to have uniform distributions within their error limits.
By multiplying these three distributions and with the appropriate constants from equation 1 we obtainedÊ
the distributions of $\dot{M}_{\rm i}$ and $\epsilon_{\rm k}$.
We finally determined the means of the distributions of 
$\dot{M}_{\rm i}$ and $\epsilon_{\rm k}$ and estimated the 68\% confidence ranges.

We note that the last stable orbit can be even smaller in a Kerr black hole
and the disk can extend down to the
event horizon. This may also be possible even for a Schwarzschild
black hole, as argued recently by Agol \& Krolik (2000).
However, at radii smaller than ~3 $R_{\rm s}$ 
the gravitational redshift would begin to become important with 
E$_{obs}$/E$_{emis}$ $\sim$ $\sqrt{(1 - 2M_{\rm bh}/r_{\rm emis})}$ 
and one would have to invoke even larger outflow velocities
to explain the detected blue-shifted Fe lines.


The total hydrogen column 
densities of the {Fe~XXV K$\alpha$} components
$abs1$ and $abs2$ are consistent with the ones determined in Chartas et al. (2003).
The mass-outflow rates and efficiencies of the outflowing components are 
listed in Table 4. 

There are several systematic errors in estimating the radii and column densities of the 
absorbers that are difficult to constrain and have not been included in the present calculations of the outflow efficiencies.
However, both these systematic uncertainties are one sided in the sense that they
lead to underestimates of the outflow efficiencies.
Specifically, we have allowed the radii of the absorbers to range between the last stable orbit of the black hole to
15 $R_{\rm s}$ based on the maximum inferred outflow velocities. 
We therefore expect any errors in the estimates of the radii of the absorbers to lead to larger values of outflow efficiencies.
Possible saturation of the absorption lines will lead to an underestimate of the column densities
and therefore to an underestimate of the outflow efficiencies as well.

\subsection{Contamination from unresolved lensed images}

The lensed images of \pgone\  
cannot be resolved with \xmm, and therefore
the spectra used in our analyses contain contributions from all images.
We estimate how this effect influences our previous results
by first determining X-ray flux ratios based on recent \chandra\ observations
of \pgone\ 
that do resolve the lensed images.
To estimate the X-ray flux ratios of \pgone\  we modeled the \chandra\ images of 
A1, A2,  B and C, with 
point-spread functions (PSFs) generated by the simulation tool \verb+MARX+ (Wise et al 1997).
The X-ray event locations were binned with a bin-size of 0\sarc0246 to sample the PSF 
sufficiently (an ACIS pixel subtends 0\sarc491).
The simulated PSFs were fitted to the \chandra\ data by minimizing the
$C$-statistic formed between the observed and simulated images
of \pgone. The relative positions of the images were fixed to
the observed NICMOS values taken from the CfA-Arizona Space Telescope Lens Survey (CASTLES).
In Figure 5 we show the Lucy-Richardson  deconvolved image of the \chandra\ observation of 
\pgone.
We find that the relative X-ray flux ratios in the 0.2--8~keV band 
are [A1/C]$_{\rm full}$ = 3.86 $\pm$ 0.42, [A2/C]$_{\rm full}$ = 1.85$\pm$ 0.22, 
and [B/C]$_{\rm full}$ = 0.93 $\pm$ 0.14.
For comparison, the {\sl HST} {\it H}-band flux ratios  
are [A1/C]$_{\rm H}$ = 4.06 $\pm$ 0.17, [A2/C]$_{\rm H}$ = 2.56 $\pm$ 0.12, 
and, [B/C]$_{\rm H}$ = 0.65 $\pm$ 0.04. The {\it H}-band magnitudes
were taken from the CASTLES WWW site.\footnote{The CASTLES WWW site is located at http://cfa-www.harvard.edu/glensdata/.}
The X-ray flux ratios of \pgone\ indicate that about 75\% of the X-ray flux originates
from the close-separation images A1 and A2. Since the observed time delay
between images A1 and A2 is $\Delta$$t_{\rm A1A2}$ = 0.149 $\pm$ 0.006~d 
there may be variability of the X-ray BALs within this time-scale only if the absorbers are launched 
near the last stable orbit of the black hole.
The ``long" time delays between images were presented in Schechter et al. (1997)  
and have image C leading images A1+A2 by 9.4 $\pm$ 3.4~d and image C leading image B by  
23.7 $\pm$ 3.4~d.
Since images B and C are considerably weaker than A1+A2
they are unlikely to be able to  
produce a significant change in the combined spectrum of \pgone.

\subsection{Comparison with \apm}

BAL quasars are relatively weak in X-rays making the detection of broad absorption 
features difficult with \xmm\ and \chandra. 
We have previously presented evidence for a relativistic outflow in 
the gravitationally lensed $z$=3.91 BAL quasar APM 08279+5255. 
Following an approach similar to the one described in \S 3.1
and using the derived outflow properties presented in Chartas et al. (2003),
we find that the mass-outflow rates and efficiencies of the outflows associated with the two Fe absorbers in \apm\ are
$\dot{M}_{\rm abs1} = 1.7_{-1.0}^{+1.4}~M_{\odot}$ yr$^{-1}$,
$\epsilon_{\rm k,abs1} = 10_{-6}^{+8} \times 10^{-3}$, 
$\dot{M}_{\rm abs2} = 3.3_{-2.1}^{+2.9}~M_{\odot}$ yr$^{-1}$, and
$\epsilon_{\rm k,abs2} = 8_{-5}^{+7} \times 10^{-2}$.
The total efficiency of the outflow will be the sum of all 
components of all species over time. 
Because of the low S/N of the available spectra of \apm\ 
additional components cannot be constrained, and we
should consider the estimated total mass-outflow rates 
and efficiencies as lower limits.
The estimated outflow rate is an order of magnitude smaller than the accretion rate in APM 08279+5255 which we estimate
to be 1.8 $\times$ 10$^{-3}$($L_{44}$/$\eta$)$M_{\odot}$  yr$^{-1}$ $\sim$ 40 $M_{\odot}$ yr$^{-1}$,
where we assumed a typical accretion efficiency of $\eta$ = 0.1.

\section{CONCLUSIONS}

High-energy X-ray absorption is detected in three epochs of monitoring observations
of the mini-BAL quasar \pgone. We interpret this absorption as due to lines arising from highly blueshifted 
Fe XXV 1s--2p and/or Fe XXVI Ly$\alpha$. We find that the depths of the X-ray
broad absorption features decreased significantly between 
epochs 1 and 2 separated by 0.92~yr (proper-time) and detect a marginal 
decrease between epochs 2 and 3 separated by 5.9 days (proper-time).
The case for a relativistic flow on the basis of either the epoch 2 or 3 observations 
of \pgone\ alone is less compelling. 
We note that models that included high-energy absorption
lines in \pgone\ and \apm\ resulted in improvements in the 
fits compared to models that did not include absorption lines at the greater than 99.9\% confidence level.
We had previously reported rapid variability over timescales of 1.8 weeks (proper-time) 
of X-ray BALs in the quasar \apm. Variability of similar magnitude and 
over similar time-scales has never been observed in UV BALs.

Assuming the interpretation that the 
high-energy absorption features are due to lines arising from highly ionized Fe XXV 
we used the measured outflow velocities, column densities, and estimated launching radii
to constrain the mass-outflow rates and outflow
efficiencies for \pgone\ and \apm. 
We find the outflow efficiencies to be 
$\epsilon_{\rm k} = 0.64_{-0.40}^{+0.52}$(68\% confidence), and 
$\epsilon_{\rm k} =0.09_{-0.05}^{+0.07}$(68\% confidence), respectively.
These estimates include only contributions from observed components 
and therefore should be considered as lower limits. 
Our derived estimates of the efficiency of the outflows in mini-BAL quasar \pgone\ and 
BAL quasar \apm,
when compared to values predicted by recent models of structure formation (SO04; G04; SDH05),
imply that these winds will have a significant impact on shaping the evolution of their host galaxies
and in regulating the growth of the central black hole.

We note that our reported values of the mass-outflow rates and outflow efficiencies 
represent instantaneous quantities. Because of the detected variability of 
the outflows in \pgone\ and \apm\ it is appropriate
to compare the outflow efficiency incorporated in theoretical models
with the average of this observed quantity over a period that covers the range of variability.
Additional monitoring observations are needed to constrain better the variability 
of the outflow properties.

The fraction of BAL quasars with highly blueshifted Fe absorption lines is not well constrained 
from available X-ray observations because of the 
relatively poor-to-moderate S/N X-ray spectra of BAL quasars observed to date.
We have shown that the two X-ray brightest mini-BAL and BAL quasars, \apm\ and \pgone, show
significant highly ionized and blueshifted Fe lines.
The third brightest BAL quasar known to date 
CSO~755 shows a hint of an Fe absorption feature in the rest-frame energy range
7.1--9.3~keV (Shemmer et al. 2005). Higher quality observations are required to confirm this result.
The X-ray spectra of the remaining observed BAL quasars are of lower S/N and
do not provide interesting constraints on the presence of high-energy Fe absorption features.  

To date, the most interesting constraints on quasar outflows are derived
from X-ray and UV spectroscopic analyses.
Current observations of BAL quasars imply that the X-ray and UV bands may be 
sampling different parts of the absorbing outflows.
A large fraction of the high velocity BAL outflow is apparently 
visible only in the X-ray band.
Unfortunately, the limited number and quality of available X-ray spectra 
leave large uncertainties in
the mass-outflow rate.  Absorption studies of quasars such as \pgone\ and \apm\ with
future high resolution, high throughput X-ray spectroscopic missions such
as {\it Constellation-X} hold great promise for pinning down this crucial
quantity.

We thank Michael Eracleous for fruitful discussions related to AGN outflows.
We thank Ohad Shemmer for pointing us to the luminosity-dependent 
bolometric correction. We greatly appreciate the useful comments made by the referee
that led to improved estimates of the black-hole mass and bolometric luminosity of PG~1115+080.
GC acknowledges financial support from NASA grants NNG04GK83G  and NAG5-13543.
WNB acknowledges financial support from NASA LTSA grant NAG5-13035. Support for SCG was provided by NASA
through the {\em Spitzer} Fellowship Program, under award 1256317.
DP acknowledges support provided by NASA through grants
HST-AR-10305.05-A and HST-AR-10680.05-A from the Space Telescope
Science Institute, which is operated by the Association of Universities
for Research in Astronomy, Inc., under NASA contract NAS5-26555.

\clearpage



\clearpage
\begin{table}
\caption{Log of Observations of BAL Quasar \pgone}
\scriptsize
\begin{center}
\begin{tabular}{lccccc}
 & & & & &\\ \hline\hline
                  &                      &                     & Exposure & Effective  &      \\
Observation & Observatory    &  Observation  &   Time  & Exposure Time\tablenotemark{a}  & $R_{sc}$\tablenotemark{b}    \\
Date           &                      &  ID                &   (ks)  & (ks)     &     \\
\hline
\hline
2004 June 26         & {\it XMM-Newton}  & 0145750101 & 86.52 & 54.92  & 0.181 $\pm$ 0.002  \\
2004 June 10         & {\it XMM-Newton}  & 0203560201 & 86.65 & 51.46  & 0.212 $\pm$ 0.002  \\
2001 November 25 &{\it XMM-Newton}  &  0082340101 & 63.36     & 52.37  & 0.202 $\pm$ 0.002   \\
2000 November 3   & {\it Chandra}          & 1630      & 9.95 & 9.83 & 0.084 $\pm$ 0.003   \\
2000 June 2           & {\it Chandra}           & 362 & 26.83 & 26.5  & 0.075 $\pm$ 0.002 \\
\hline \hline
\end{tabular}
\end{center}
\tablenotetext{a}{Effective exposure time is the time remaining after the application of good time-interval (GTI)
tables to remove portions of the observation that were severely contaminated by background.}
\tablenotetext{b}{Background-subtracted source count rate including events with energies within the 0.2--10~keV band.
The source count rates and effective exposure times for the \xmm\ observations refer to those obtained with the
EPIC PN instrument.
Please see \S 2 for details on source and background
extraction regions used for measuring $R_{sc}$.}
\end{table}

\clearpage
\begin{table}
\caption{RESULTS FROM FITS TO THE \xmm\ EPIC SPECTRA OF \pgone}
\scriptsize
\begin{center}
\begin{tabular}{ccccc}
 & & & & \\ \hline\hline
\multicolumn{1}{c} {Model$^{a}$} &
\multicolumn{1}{c} {Parameter$^{b}$} &
\multicolumn{1}{c} {Values For Epoch 1$^{c}$} &
\multicolumn{1}{c} {Values For Epoch 2$^{c}$} &
\multicolumn{1}{c} {Values For Epoch 3$^{c}$} \\
  &              & 2001 November 25 &               2004 June 10                                &     2004 June 26                            \\
    &              &&                                              &                                \\
1 &$\Gamma$     &  1.85$_{-0.03}^{+0.03}$ & 1.85$_{-0.03}^{+0.03}$ &   1.80$_{-0.03}^{+0.03}$      \\
&   $N_{\rm H}$ & (0.22$_{-0.05}^{+0.05}$) $\times$ 10$^{22}$~cm$^{-2}$   & (0.46$_{-0.05}^{+0.05}$) $\times$ 10$^{22}$~cm$^{-2}$ & (0.37$_{-0.05}^{+0.05}$) $\times$ 10$^{22}$~cm$^{-2}$ \\
 &$\chi^2/{\nu}$ & 361.2/240 & 361.1/272 & 344.8/255 \\
&$P(\chi^2/{\nu})$$^{d}$ & 6.6~$\times$~10$^{-7}$  &2.3~$\times$~10$^{-4}$&  1.5~$\times$~10$^{-4}$   \\
  &                                                             &           &           &        \\
  &                                         &                             &                          &                \\
${2}$ &  $\Gamma$ & 1.68$_{-0.04}^{+0.04}$ &1.83$_{-0.03}^{+0.03}$&1.72$_{-0.03}^{+0.03}$                           \\
  &   $N_{\rm H}$ & (0.12$_{-0.03}^{+0.03}$) $\times$ 10$^{22}$~cm$^{-2}$  &(0.44$_{-0.05}^{+0.05}$) $\times$ 10$^{22}$~cm$^{-2}$ &(0.32$_{-0.05}^{+0.05}$) $\times$ 10$^{22}$~cm$^{-2}$ \\
  &  E$_{\rm abs1}$        & 7.27$_{-0.10}^{+0.37}$~keV           &6.40$_{-0.16}^{+0.14}$~keV& $---$   \\
  &  $\sigma_{\rm abs1}$ & $ < $ 0.50~keV        &  $ < $ 0.44~keV & $---$      \\
  &  EW$_{\rm abs1}$     & 0.117$_{-0.077}^{+0.089}$~keV          &0.057$_{-0.057}^{+0.072}$~keV&  $---$        \\
  &  E$_{\rm abs2}$        & 10.51$_{-1.25}^{+0.79}$~keV           &8.53$_{-0.08}^{+0.30}$~keV& $9.6_{-0.76}^{+1.7}$~keV   \\
  &  $\sigma_{\rm abs2}$ & 3.49$_{-0.83}^{+1.82}$~keV        & $ < $ 0.53 &  $2.7_{-0.7}^{+2.1}$~keV      \\
  &  EW$_{\rm abs2}$     & 2.27$_{-0.47}^{+0.46}$~keV          & $0.18_{-0.10}^{+0.11}$~keV &   $1.3_{-0.4}^{+0.7}$~keV         \\
  & $\chi^2/{\nu}$     & 284.8/228                        &320.8/260 &  363.6/250      \\
  & $P(\chi^2/{\nu})$$^{d}$  & 6.27~$\times$~10$^{-3}$           & 6.04~$\times$~10$^{-3}$ &     3.47~$\times$~10$^{-6}$                  \\
  &           &                    &                   &                                          \\
${3}$ &$\Gamma$ & 1.84$_{-0.03}^{+0.05}$ &1.92$_{-0.04}^{+0.04}$& $1.88_{-0.04}^{+0.04}$                     \\
  &   $N_{\rm H}$ & (1.33$_{-0.35}^{+0.58}$) $\times$ 10$^{22}$~cm$^{-2}$   &(1.41$_{-0.27}^{+0.36}$) $\times$ 10$^{22}$~cm$^{-2}$ &(1.52$_{-0.37}^{+0.46}$) $\times$ 10$^{22}$~cm$^{-2}$\\
  &    $\xi$            & 144.9$_{-52.3}^{+83.6}$~erg~cm~s$^{-1}$   &55.9$_{-23.4}^{+29.6}$~erg~cm~s$^{-1}$&90.2$_{-38.5}^{+43.3}$~erg~cm~s$^{-1}$  \\
  & E$_{\rm abs1}$       & 7.27$_{-0.13}^{+0.20}$~keV       &6.40$_{-0.13}^{+0.13}$~keV & $---$    \\
  &  $\sigma_{\rm abs1}$&$ < $ 0.44~keV     & $ < $ 0.33~keV &  $---$ \\
  &  EW$_{\rm abs1}$    & $0.113_{-0.077}^{+0.093}$~keV        &$0.069_{-0.069}^{+0.073}$~keV & $---$          \\
  & E$_{\rm abs2}$        & 9.79$_{-1.05}^{+0.96}$~keV       &8.61$_{-0.17}^{+0.23}$~keV&$8.9_{-0.46}^{+0.73}$~keV (68\% level)      \\
  &  $\sigma_{\rm abs2}$& 3.07$_{-0.83}^{+1.82}$~keV    & $ < $ 0.51~keV&$0.5_{-0.24}^{+1.7}$~keV  (68\% level) \\
  &  EW$_{\rm abs2}$    & $2.05_{-0.47}^{+0.52}$~keV        &$0.17_{-0.10}^{+0.11}$~keV&   $0.28_{-0.11}^{+0.39}$~keV (68\% level)         \\
  &   $\chi^2/{\nu}$   & 241.5/227                     &265.1/259 & 278.8/249         \\
  &  $P(\chi^2/{\nu})$$^{d}$ & 0.20                   &0.38  & 0.10           \\
  &  &&                   &                                          \\
  ${4}$ &$\Gamma$ & 1.84$_{-0.03}^{+0.05}$ &1.94$_{-0.04}^{+0.02}$& $1.90_{-0.04}^{+0.04}$                     \\
  &   $N_{\rm H}$ & (1.97$_{-0.49}^{+0.70}$) $\times$ 10$^{22}$~cm$^{-2}$   &(1.61$_{-0.30}^{+0.30}$) $\times$ 10$^{22}$~cm$^{-2}$ &(1.73$_{-0.37}^{+0.39}$) $\times$ 10$^{22}$~cm$^{-2}$\\
  &    $f$            & 0.40$_{-0.05}^{+0.06}$   &0.65$_{-0.04}^{+0.03}$&0.57$_{-0.04}^{+0.04}$  \\
  & E$_{\rm abs1}$       & 7.27$_{-0.13}^{+0.28}$~keV       &6.41$_{-0.13}^{+0.13}$~keV & $---$    \\
  &  $\sigma_{\rm abs1}$&$ < $ 0.44~keV     & $ < $ 0.33~keV &  $---$ \\
  &  EW$_{\rm abs1}$    & $0.103_{-0.072}^{+0.082}$~keV        &$0.075_{-0.075}^{+0.075}$~keV & $---$          \\
  & E$_{\rm abs2}$        & 9.62$_{-1.05}^{+0.86}$~keV       &8.53$_{-0.08}^{+0.26}$~keV&$8.9_{-0.3}^{+0.5}$~keV (68\% level)      \\
  &  $\sigma_{\rm abs2}$& 3.18$_{-0.71}^{+0.95}$~keV    & $ < $ 0.44~keV&$0.5_{-0.2}^{+1.3}$~keV  (68\% level) \\
  &  EW$_{\rm abs2}$    & $2.01_{-0.48}^{+0.55}$~keV        &$0.19_{-0.11}^{+0.11}$~keV&   $0.28_{-0.13}^{+0.30}$~keV (68\% level)         \\
  &   $\chi^2/{\nu}$   & 247.6/227                     &263.9/259 & 279.6/249         \\
  &  $P(\chi^2/{\nu})$$^{d}$ & 0.17                   &0.40  & 0.09           \\
\hline \hline
\end{tabular}
\end{center}
\noindent
${}^{a}$ Model 1 consists of a power law and neutral absorption at the source. 
Model 2  consists of a power law, neutral absorption at the source, and Gaussian absorption lines at the source.
Model 3  consists of a power law, ionized absorption at the source, and Gaussian absorption lines at the source.
Model 4 consists of a power law, partial covering of the source, and Gaussian absorption lines at the source.
All model fits include the Galactic absorption toward the source (Stark et al. 1992).
All fits have been performed on the combined spectrum of images A1, A2, B and C of PG~1115+080. \\
${}^{b}$All absorption-line parameters are calculated for the rest frame.\\
${}^{c}$All errors are for 90\% confidence unless mentioned otherwise with all
parameters taken to be of interest except absolute normalization.\\
${}^{d}$$P(\chi^2/{\nu})$ is the probability of exceeding $\chi^{2}$ for ${\nu}$ degrees of freedom
if the model is correct.
\end{table}
\clearpage
\begin{table}
\caption{Energies and Inferred Outflow Velocities of Absorption Lines Detected in the \xmm\ EPIC Observations of \pgone\ During Epoch 1}
\scriptsize
\begin{center}
\begin{tabular}{lccrrllc}
 & & & & & & &\\ \hline\hline
\multicolumn{1}{c} {Line $^{a}$} &
\multicolumn{1}{c} {Component} &
\multicolumn{1}{c} {$E_{\rm obs}$} &
\multicolumn{1}{c} {$E_{\rm rest}$} &
\multicolumn{1}{c} {$E_{\rm lab}$} &
\multicolumn{1}{c} {$EW_{\rm rest}$} &
\multicolumn{1}{c} {$\sigma_{\rm rest}$} &
\multicolumn{1}{c} {$v_{\rm abs}$} \\
\multicolumn{1}{c} {} &
\multicolumn{1}{c} {} &
\multicolumn{1}{c} {(keV)} &
\multicolumn{1}{c} {(keV)} &
\multicolumn{1}{c} {(keV)} &
\multicolumn{1}{c} {(keV)} &
\multicolumn{1}{c} {(keV)} &
\multicolumn{1}{c} {$(c)$} \\
                                  &        &                 & &                          &                                       &                   &           \\
  Fe XXV 1s--2p  & $abs1$  &  $2.67_{-0.04}^{+0.14}$    & $7.27_{-0.10}^{+0.37}$  &  6.7       & 0.11$_{-0.08}^{+0.09}$ & $ < $0.44 &   $0.09_{-0.02}^{+0.05}$     \\
  Fe XXV 1s--2p & $abs2$  &  $3.60_{-0.46}^{+0.29}$    & $9.79_{-1.05}^{+0.96}$ &  6.7       & 2.05$_{-0.47}^{+0.52}$ &3.07$_{-0.83}^{+1.82}$& $0.40_{-0.11}^{+0.09}$ \\
\hline \hline
\end{tabular}
\end{center}
\tablecomments{The present spectra cannot distinguish whether the absorption in components $abs1$ and $abs2$ is caused by
Fe~XXV~1s--2p or Fe~XXVI~Ly${\alpha}$.}
\end{table}

\clearpage 
\begin{table}
\caption{Hydrogen Column Densities, Mass-Outflow Rates and Efficiencies of Outflows in PG~1115+080 and APM~08279+0552}
\scriptsize
\begin{center}
\begin{tabular}{ccccc}
 & & & & \\ \hline\hline
 Component & Line & N$_{\rm H}$& $\dot{M}$ & $\epsilon_{\rm k}$  \\
                 &          & cm$^{-2}$  &(M$_{\odot}~yr^{-1}$)  &       \\
&    &    &    &  \\
\multicolumn{5}{c}{\pgone\ Outflow} \\
\hline
&    &    &   &   \\
$abs1$         & Fe XXV 1s--2p  & 4 $\times$ 10$^{22}$ & $0.10_{-0.07}^{+0.09} $        & $7.2_{-4.6}^{+6.3} \times 10^{-4}$ \\
$abs2$         & Fe XXV 1s--2p  & 4 $\times$ 10$^{23}$ & $4.6_{-2.9}^{+4.0}$                         & $6.3_{-4.0}^{+5.5} \times 10^{-1}$ \\
\hline
&    &    &    & \\
\multicolumn{5}{c}{\apm\ Outflow} \\
\hline
&    &    &      &\\
$abs1$         &Fe XXV 1s--2p  & 1 $\times$ 10$^{23}$ &$1.7_{-1.0}^{+1.4}$  & $1.0_{-0.6}^{+0.8} \times 10^{-2}$   \\
$abs2$         & Fe XXV 1s--2p & 1 $\times$ 10$^{23}$ &$3.3_{-2.1}^{+2.9}$ &  $0.8_{-0.5}^{+0.7} \times 10^{-1}$   \\
\hline \hline
\end{tabular}
\end{center}
\tablecomments{The bolometric luminosites of PG~1115+080 and APM~08279+0552  are 3.3~$\times$~10$^{46}$~erg~s$^{-1}$ 
and 2~$\times$~10$^{47}$~erg~s$^{-1}$, respectively.}
\end{table}

\clearpage

\begin{figure*}
\centerline{\includegraphics[width=14cm]{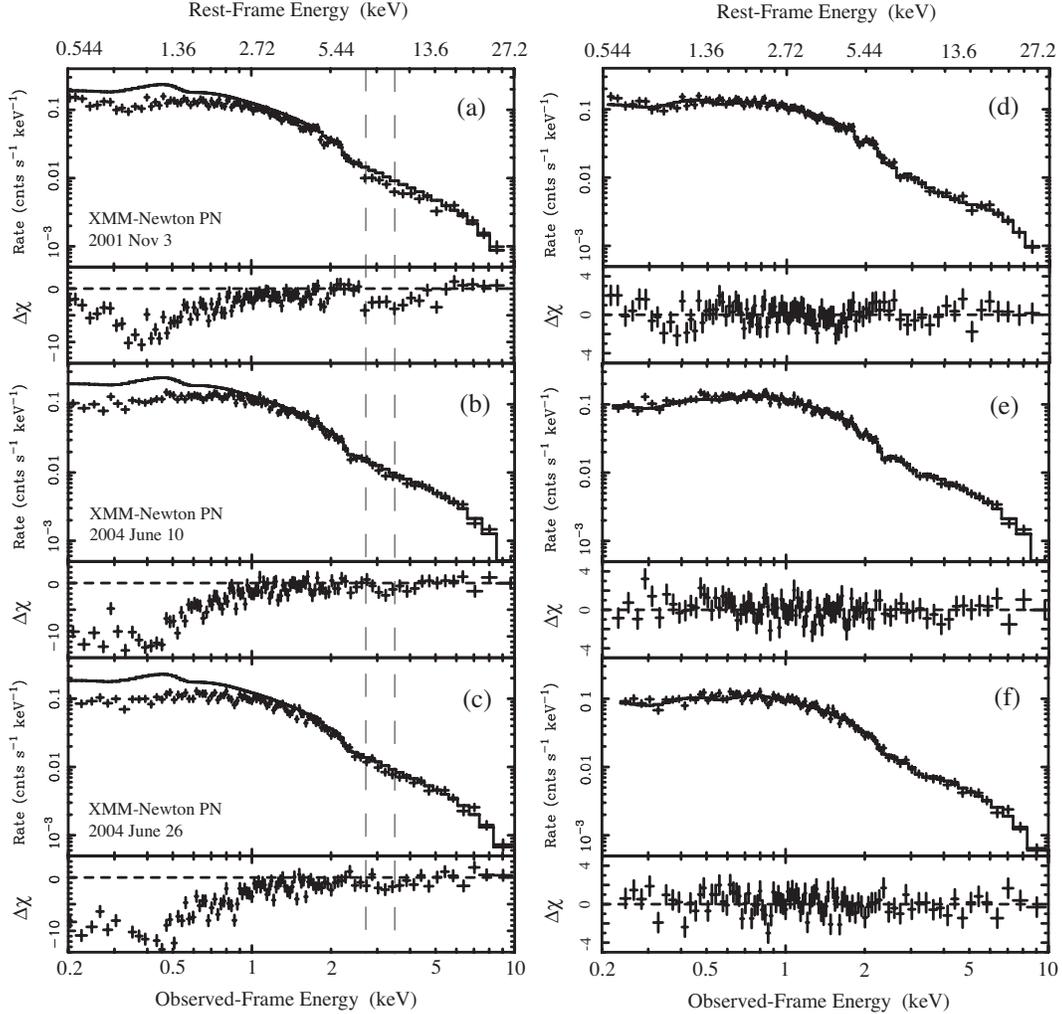}}
\caption{(a) The top panel shows the \xmm\ PN spectrum of
the combined images of \pgone\ for Epoch 1 fit with Galactic absorption and a 
power-law model to events with energies lying within 
the observed-frame ranges of 2--2.5~keV and 5--10~keV. 
The best-fit values of the power-law photon indices in these energy ranges for the three epochs were almost identical  
with $\Gamma$ $\sim$ 1.9.  
The lower panel shows the residuals of the fit in units of 1$\sigma$ deviations.
Several absorption features within the observed-frame range of 1.5--5.2~keV are noticeable in the residuals plot. 
(b) Same as (a) for Epoch 2.
(c) Same as (a) for Epoch 3.
For clarity we only show the higher S/N ratio PN data; however, all fits were performed 
simultaneously using the PN and MOS1+2 data unless mentioned otherwise.
The vertical dashed lines indicate the best-fit energies of the high-energy absorption lines for epoch 1.
The top panels (d), (e) , and (f) show the same data shown in panels (a), (b), and (c) overplotted with the best-fit models
taken from \hbox{model 3} of \hbox{Table 2}. The lower panels of (d), (e), and (f) show the reduced residuals 
of these fits.
\label{fig1.eps}}
\end{figure*}

\clearpage
\begin{figure*}
\centerline{\includegraphics[width=14cm]{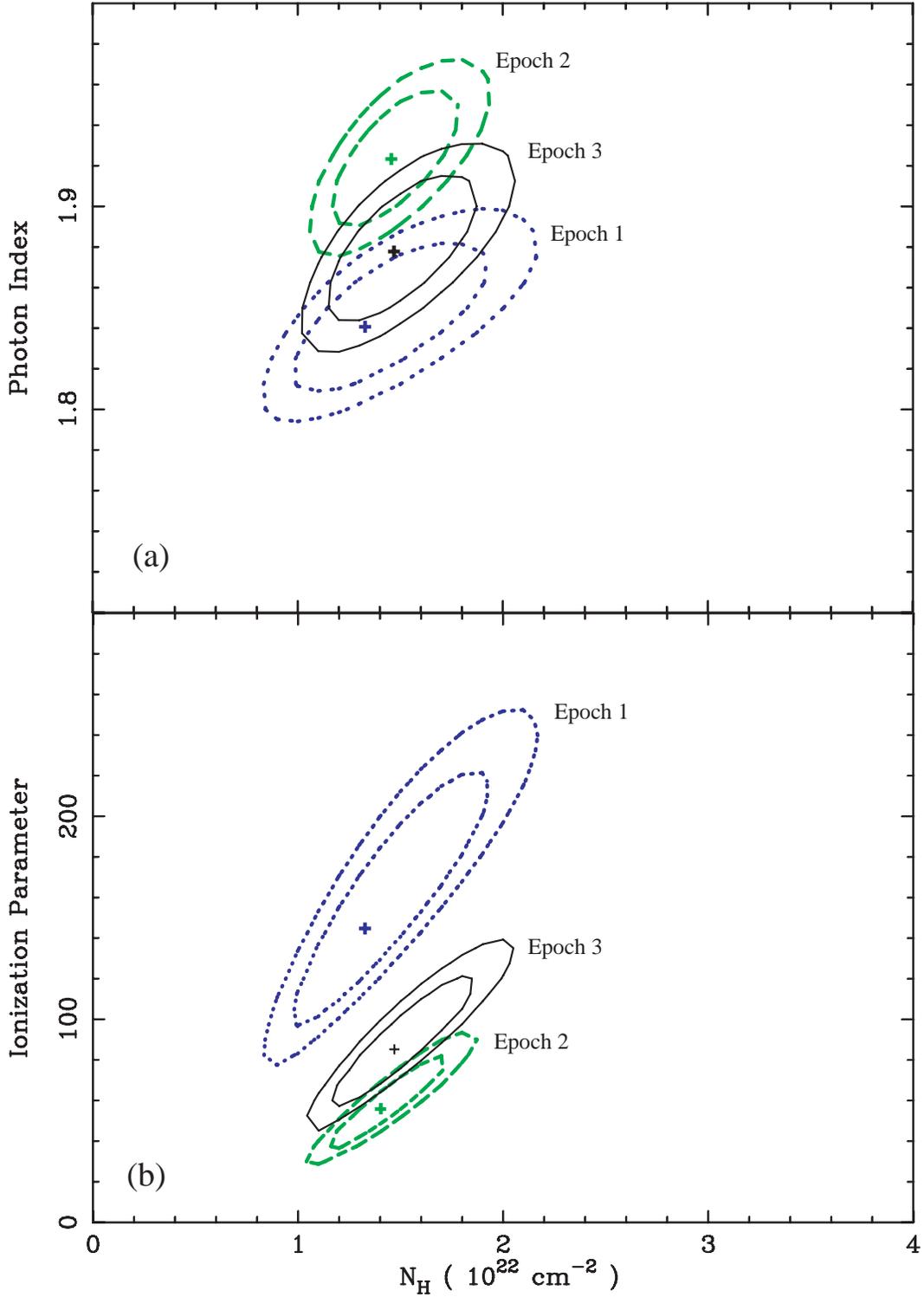}}
\caption{68\% and 90\%  confidence contours between 
(a) the photon index 
and the intrinsic-absorption column density,
and (b) the ionization parameter and the intrinsic-absorption column density
for the \xmm\ observations of \pgone\ during epoch 1 (blue dotted contours),
epoch 2 (green dashed contours) and epoch 3 (black solid contours) 
assuming Model 3 of Table 2. The ionization parameter and column density are for the
low-energy absorber.
\label{fig2.eps}}
\end{figure*}

\clearpage

\begin{figure*}
\centerline{\includegraphics[width=14cm]{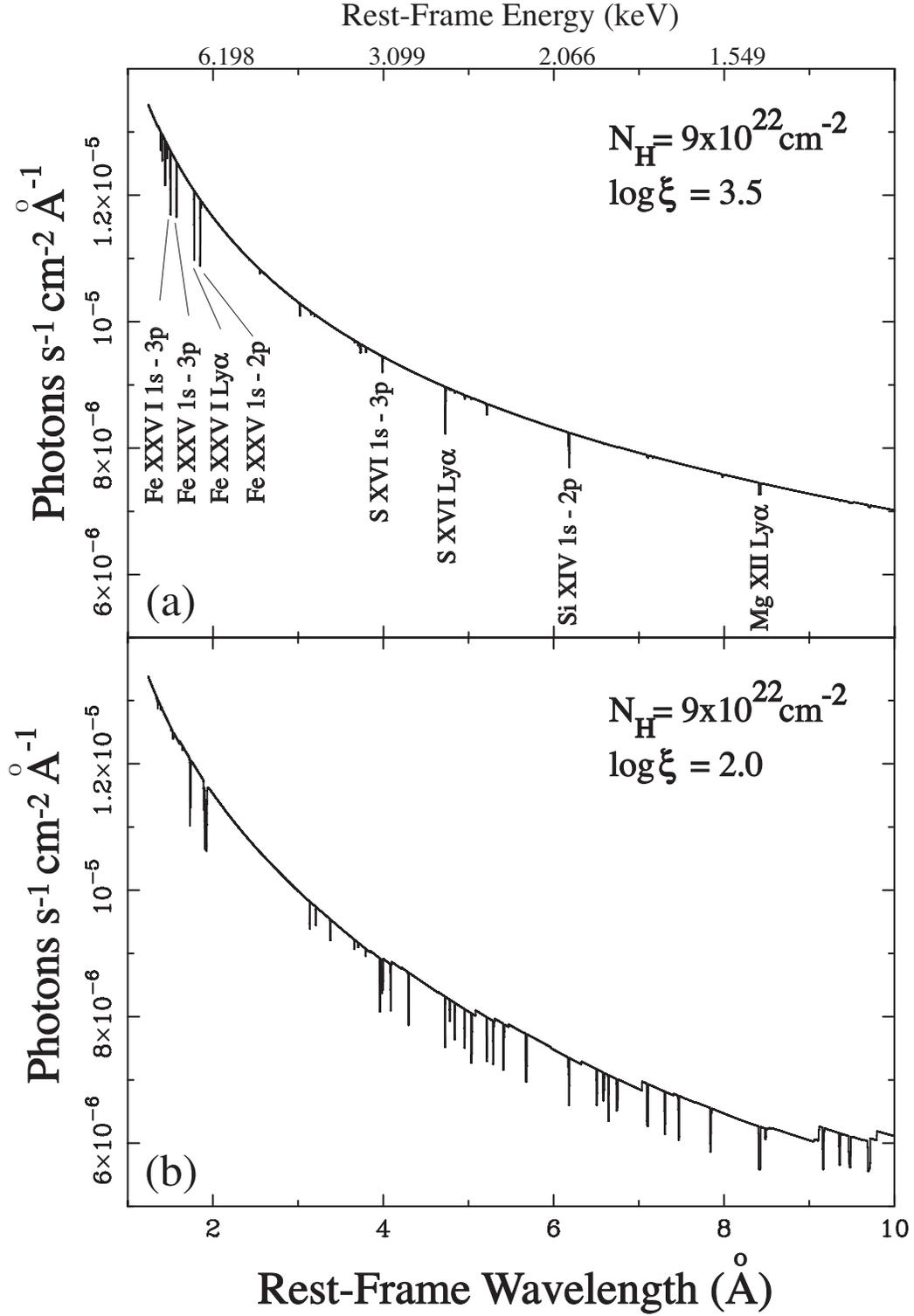}}
\caption{Simulations of absorbed spectra with the continuum level
normalized to that of \pgone.  In panels a and b we show the absorbed spectra for
ionization parameters of $\log\xi$= 3.5 and $\log\xi$=2.0, respectively. 
\label{fig3.eps}}
\end{figure*}

\clearpage

\begin{figure*}
\centerline{\includegraphics[width=14cm]{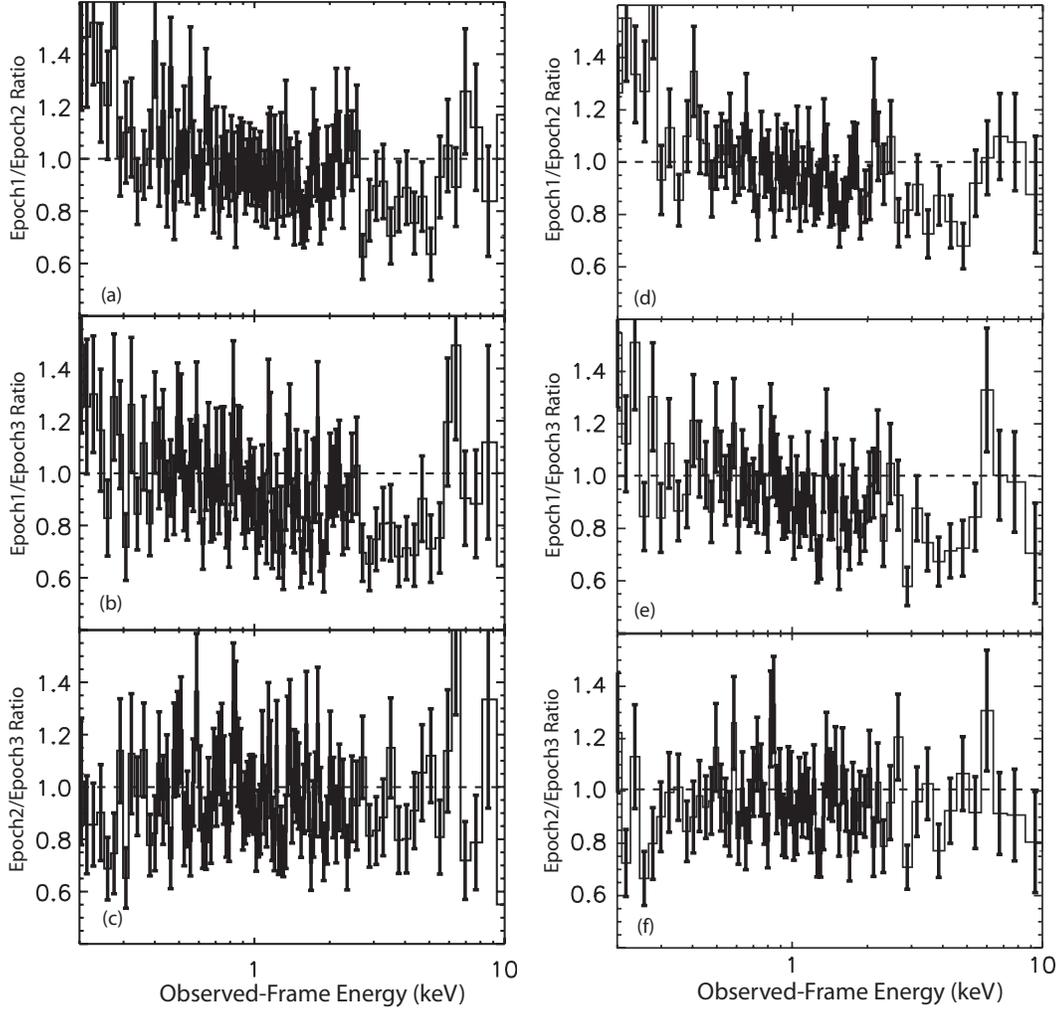}}
\caption{The ratio of the \pgone\ spectra of (a) epoch 1 to epoch 2, (b) epoch 1 to {epoch~3} and, 
(c) epoch 2 to {epoch~3}. The spectra are binned with the same grouping used for epoch 1.
The spectrum of epoch 1 was binned such that the minimum counts per bin was 100 counts.
Panels (d), (e), and (f)  on the right show the same ratios as the panels on the left with the only difference being 
that the spectrum of epoch 1 was binned such that the minimum counts per bin was 200 counts.
We note that the PN background for these observations becomes increasingly 
significant compared to the source signal for observed-frame energies above 6.5~keV.
This explains the larger error bars and dispersion of the data points for the ratios above 6.5~keV.
 \label{fig4.eps}}
\end{figure*}

\clearpage

\begin{figure*}
\centerline{\includegraphics[width=14cm]{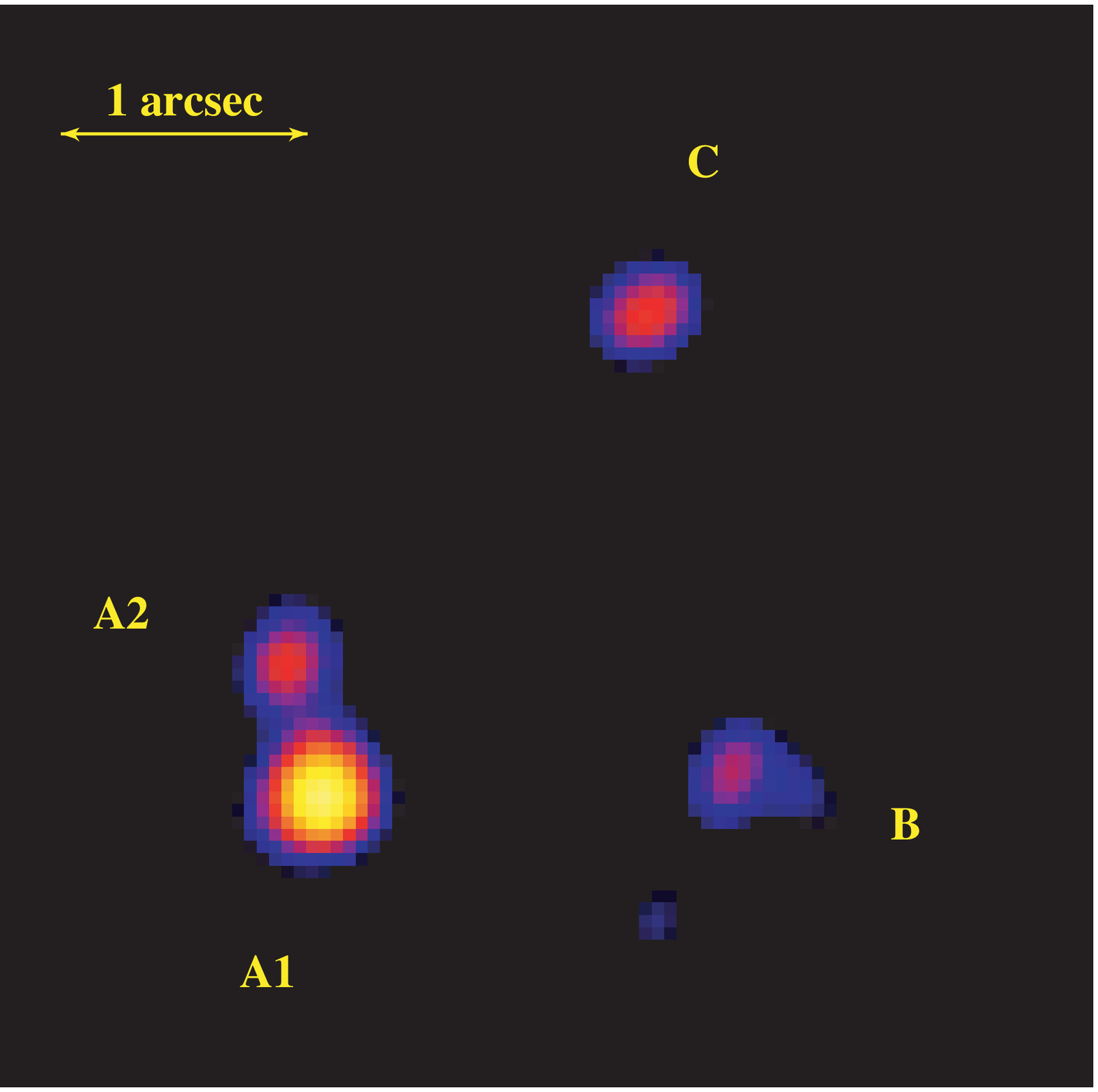}}
\caption{ 
Lucy-Richardson  deconvolved image (0.2 -- 8~keV band) of the 
\chandra\ observation of PG~1115+080.
 \label{fig5.eps}}
\end{figure*}


\begin{references}

\reference{} Agol, E., \& Krolik, J.~H.\ 2000, \apj, 528, 161 \\

\reference{} Aldcroft, T.~L., \& Green, P.~J.\ 2003, \apj, 592, 710 \\


\reference{} Arav, N., Li, Z., \& Begelman, M.~C.\ 1994, \apj, 432, 62 \\

\reference{} Arav, N., Barlow, T.~A., 
Laor, A., Sargent, W.~L.~W., \& Blandford, R.~D.\ 1998, \mnras, 297, 990 \\


\reference{} Arnaud, K.~A.\ 1996, ASP 
Conf.~Ser.~101: Astronomical Data Analysis Software and Systems V, 5, 17 \\


\reference{} Barlow, T.~A., Hamann, 
F., \& Sargent, W.~L.~W.\ 1997, ASP Conf.~Ser.~128: Mass Ejection from 
Active Galactic Nuclei, 128, 13 \\


\reference{} {Braito}, V., {Della Ceca}, R., {Piconcelli}, E., et al.,
2004, \aap, 420, 79 \\


\reference{} Chartas, G.\ 2000, \apj, 531, 81 \\

\reference{} Chartas, G., Brandt, W.~N., Gallagher, S.~C., \& Garmire, G.~P.\ 2002, \apj, 579, 169 \\

\reference{} Chartas, G., Brandt, W.~N., \& Gallagher, S.~C.\ 2003, ApJ, 595, 85 \\


\reference{} Crenshaw, D.~M., 
Kraemer, S.~B., Boggess, A., Maran, S.~P., Mushotzky, R.~F., \& Wu, C.\ 
1999, ApJ, 516, 750 \\

\reference{} Crenshaw, D.~M., Kraemer, S.~B., Gabel, J.~R., et al.,
2003, ApJ, 594, 116 \\


\reference{} Done, C., Mulchaey, J.~S., Mushotzky, R.~F., \& Arnaud, K.~A.\ 1992, \apj, 395, 275 \\


\reference{} {Elvis}, M., {Wilkes}, B.~J., {McDowell}, J.~C., {Green}, R.~F., 
{Bechtold}, J., {Willner}, S.~P., {Oey}, M.~S., {Polomski}, E., \& 
{Cutri}, R., 1994, \apjs, 95, 1 \\

\reference{} Elvis, M.\ 2000, \apj, 545, 63 \\

\reference{} Everett, J.~E., \& Ballantyne, D.~R.\ 2004, \apjl, 615, L13 \\

\reference{} Everett, J.~E.\ 2005, \apj, 631, 689 \\


\reference{} Gallagher, S.~C., Brandt, W.~N., Chartas, G., \& Garmire, G.~P.\ 2002, \apj, 567, 37 \\


\reference{} Garmire, G.~P., Bautz, 
M.~W., Ford, P.~G., Nousek, J.~A., \& Ricker, G.~R.\ 2003, \procspie, 4851, 28 \\

\reference{} Goodrich, R.~W.\ 1997, \apj,  474, 606 \\

\reference{} Granato, G.~L., De 
Zotti, G., Silva, L., Bressan, A., \& Danese, L.\ 2004, \apj, 600, 580 \\


\reference{} Green, P.~J., Aldcroft, T.~L., Mathur, S., Wilkes, B.~J., \& Elvis, M.\ 2001, \apj, 558, 109 \\

\reference{} Green, R.~F., Pier, J.~R., Schmidt, M., Estabrook, F.~B., Lane, A.~L., \& Wahlquist, H.~D.\ 
1980, \apj, 239, 483 \\

\reference{} Hamann, F., Barlow, 
T.~A., Beaver, E.~A., Burbidge, E.~M., Cohen, R.~D., Junkkarinen, V., \& 
Lyons, R.\ 1995, \apj, 443, 606 \\


\reference{} Hasinger, G., Schartel, N., \& Komossa, S.\ 2002, \apjl, 573, L77 (H02) \\ 

\reference{} Hewett, P.~C., \& Foltz, C.~B.\ 2003, \aj, 125, 1784 \\

\reference{} Hopkins, P.~F., 
Hernquist, L., Cox, T.~J., Di Matteo, T., Martini, P., Robertson, B., \& 
Springel, V.\ 2005, \apj, 630, 705 \\

\reference{} Hopkins, P.~F., 
Hernquist, L., Cox, T.~J., Di Matteo, T., Robertson, B., \& Springel, V.\ 
2006, \apjs, 163, 1 \\


\reference{} Impey, C.~D., Falco, 
E.~E., Kochanek, C.~S., Leh{\' a}r, J., McLeod, B.~A., Rix, H.-W., Peng, 
C.~Y., \& Keeton, C.~R.\ 1998, \apj, 509, 551 \\

\reference{} Irwin, M.~J., Ibata, R.~A., Lewis, G.~F., \& Totten, E.~J.\ 1998, \apj, 505, 529 \\


\reference{} Jansen, F., Lumb, D., Altieri, B., et al. 2001, A\&A, 365, L1\\

\reference{} Kaastra, J.~S., 
Steenbrugge, K.~C., Raassen, A.~J.~J., van der Meer, R.~L.~J., Brinkman, 
A.~C., Liedahl, D.~A., Behar, E., \& de Rosa, A.\ 2002, \aap, 386, 427 \\


\reference{} {Kaspi}, S., {Brandt}, W.~N., {George}, I.~M., et al.,
2002, ApJ, 574, 643 \\


\reference{} Kaspi, S., \& Behar, E.\ 2006, \apj, 636, 674 \\




\reference{} Kraemer, S.~B., Crenshaw, D.~M., Hutchings, J.~B., et al.,
2001, ApJ, 551, 671 \\

\reference{} Kriss, G. A. 2002, in ASP Conf. Ser 255, Mass Outflow in
Active Galacatic Nuclei: Perspectives, ed. D. M. Crenshaw, S. B. Kraemer, \& I. M. George
(San Francisco; ASP), 69 \\


\reference{} Krolik, J.~H.~\& Voit, G.~M.\ 1998, \apjl, 497, L5 \\

\reference{} McKernan, B., Yaqoob, 
T., \& Reynolds, C.~S.\ 2004, \apj, 617, 232 \\


\reference{} McKernan, B., Yaqoob, 
T., \& Reynolds, C.~S.\ 2005, \mnras, 361, 1337 \\

\reference{} Maloney, P.~R., \& 
Reynolds, C.~S.\ 2000, \apjl, 545, L23 \\

\reference{} Mathur, S.~\& Williams, R.~J. \ 2002, \apjl, 589, L1 \\ 


\reference{} Michalitsianos, A.~G., Oliversen, R.~J., \& Nichols, J.\ 1996, \apj, 461, 593 \\

\reference{} Murray, N., Chiang, J., 
Grossman, S.~A., \& Voit, G.~M.\ 1995, \apj, 451, 498 \\

\reference{} {Netzer}, H., {Kaspi}, S., {Behar}, E., et al.,
2003, ApJ, 599, 933 \\

\reference{} Neugebauer, G., 
Green, R.~F., Matthews, K., Schmidt, M., Soifer, B.~T., \& Bennett, J.\ 
1987, \apjs, 63, 615 \\


\reference{} Porquet, D., Reeves, 
J.~N., O'Brien, P., \& Brinkmann, W.\ 2004, \aap, 422, 85 \\


\reference{} Pounds, K.~A., King, 
A.~R., Page, K.~L., \& O'Brien, P.~T.\ 2003a, \mnras, 346, 1025 \\

\reference{} Pounds, K.~A., Reeves, 
J.~N., King, A.~R., Page, K.~L., O'Brien, P.~T., \& Turner, M.~J.~L.\ 2003, 
\mnras, 345, 705 \\

\reference{} Proga, D., Stone, J.~M., \& Kallman, T.~R.\ 2000, \apj, 543, 686 \\

\reference{} Proga, D., \& 
Kallman, T.~R.\ 2004, \apj, 616, 688 \\

\reference{} Punsly, B.\ 2006, \apj, 647, 886 \\

\reference{} Reeves, J.~N., O'Brien, P.~T., \& Ward, M.~J.\ 2003, \apjl, 593, L65 \\

\reference{} Reeves, J.~N., \& Turner, M.~J.~L.\ 2000, \mnras, 316, 234 \\

\reference{} Scannapieco, E., \& Oh, S.~P.\ 2004, \apj, 608, 62 \\

\reference{} {Schechter}, P.~L., {Bailyn}, C.~D., {Barr}, R., et al.,
1997, \apjl, 475, L85 \\

\reference{} Shemmer, O., Brandt, W. N., Gallagher, S. C., Vignali, C., Boller, Th., Chartas, G., 
\& Comastri, A., \ 2005, \aj, 130, 2522 \\

\reference{} Sim, S.~A.\ 2005, \mnras, 356, 531 \\

\reference{} Spitzer, L. 1978, Physical Processes in the Interstellar Medium (New York: Wiley) \\

\reference{} Springel, V., Di 
Matteo, T., \& Hernquist, L.\ 2005, \apjl, 620, L79 \\

\reference{} Srianand, R., \& Petitjean, P.\ 2000, \aap, 357, 414 \\

\reference{} Stark, A.~A., Gammie, C.~F., Wilson, R.~W., Bally, J., 
Linke, R.~A., Heiles, C., \& Hurwitz, M.\ 1992, \apjs, 79, 77 \\

\reference{} Steffen, A.~T., 
Strateva, I., Brandt, W.~N., Alexander, D.~M., Koekemoer, A.~M., Lehmer, 
B.~D., Schneider, D.~P., \& Vignali, C.\ 2006, \aj, 131, 2826 \\

\reference{}
{Str{\" u}der}, L., {Briel}, U., {Dennerl}, K., et al.,
 \ 2001, \aap, 365, L18 \\
 
\reference{} Tananbaum, H. Avni, Y., Branduardi, G., Elvis, M., 
Fabbiano, G., Feigelson, E., Giacconi, R., Henry, J.~P., Pye, 
J.~P., Soltan, A., Zamorani, G.\ 1979, \apjl, 234, L9 \\

\reference{} Tolea, A., Krolik, J.~H., \& Tsvetanov, Z.\ 2002, \apjl, 578, L31 \\

\reference{} Tripp, T. M., Green, R. F., \& Bechtold, J.  1990, \apjl, 364, L29 \\

\reference{} {Turner}, M.~J.~L., {Abbey}, A., {Arnaud}, M., 
\ 2001, \aap, 365, L27 \\

\reference{} Tsunemi, H., Mori, K., 
Miyata, E., Baluta, C., Burrows, D.~N., Garmire, G.~P., \& Chartas, G.\ 
2001, \apj, 554, 496 \\

\reference{} Turner, T.~J., \& 
Kraemer, S.~B.\ 2003, \apj, 598, 916 \\


\reference{} Turnshek, D.~A., 
Grillmair, C.~J., Foltz, C.~B., \& Weymann, R.~J.\ 1988, \apj, 325, 651 \\

\reference{} Young, P., Sargent, 
W.~L.~W., \& Boksenberg, A.\ 1982, \apj, 252, 10 \\


\reference{} Verner, D.~A., Verner, E. M.,  and G. J. Ferland, G. J.,  1996, Atomic Data Nucl. Data Tables, 64, 1 \\

\reference{} Verner, D.~A.~\& Yakovlev, D.~G.\ 1995, \aaps, 109, 125 \\

\reference{} Vestergaard, 
M., \& Peterson, B.~M.\ 2006, \apj, 641, 689 \\


\reference{} Weymann, R.~J., Latham, D., Roger, J., Angel, P., Green, R.~F., 
Liebert, J.~W., Turnshek, D.~A., Turnshek, D.~E., and Tyson, J.~A. \ 
1980, \nat, 285, 641 \\

\reference{} Weymann, R.~J., Morris, S.~L., Foltz, C.~B., \& Hewett, P.~C.\ 1991, \apj, 
373, 23 \\

\reference{} Wise, M.~W., Huenemoerder, 
D.~P., \& Davis, J.~E.\ 1997, ASP Conf.~Ser.~125: Astronomical Data 
Analysis Software and Systems VI, 125, 477 

\end{references}
\end{document}